\documentclass[12pt]{article}
\pdfoutput=1
\usepackage[utf8]{inputenc}
\usepackage[DIV13]{typearea}
\usepackage{indentfirst}
\usepackage{calligra,amsmath,amsfonts,mathrsfs,amssymb,bbm}
\usepackage{slashed,cancel,units}
\usepackage{array,longtable,multirow,multicol,colortbl}
\usepackage{graphicx}
\usepackage[usenames,dvipsnames]{xcolor}
\usepackage{cite}
\usepackage{subfigure}
\usepackage[margin=0pt]{caption}
\usepackage{hyperref}
\hypersetup{colorlinks=true,urlcolor=Magenta,anchorcolor=blue,citecolor=blue,filecolor=blue,linkcolor=Magenta,menucolor=blue,linktocpage=true,pdfproducer=medialab}

\newcommand{\email}[1]{\href{mailto:#1}{\tt #1}}

\numberwithin{equation}{section}

\def\be{\begin{equation}}
\def\ee{\end{equation}}

\def\hc{\text{h.c.}}

\def\diag{\mathtt{diag}}
\newcommand{\vev}[1]{\langle #1\rangle}
\newcommand{\ov}[1]{\overline{#1}}
\def\unity{\mathbbm 1}

\def\g{\text{\calligra g}}
\def\G{\mathcal G}
\def\HH{\mathscr H}
\def\LL{\mathscr L}
\def\O{\mathcal O}
\def\U{\mathcal U}
\def\Y{\mathcal Y}

\def\eV{\text{ eV}}
\def\GeV{\text{ GeV}}

\def\Dmsol{\Delta m^2_\text{sol}}
\def\Dmatm{\Delta m^2_\text{atm}}
\def\LLF{\Lambda_\text{LFV}}
\def\LLN{\Lambda_\text{L}}
\def\muLN{\mu_\text{L}}

\newcommand{\blue}[1]{\color{blue} #1 \color{black}}
\begin{document}
\begin{titlepage}
\vspace*{-1cm}
\phantom{hep-ph/***}
\flushleft{\blue{FTUAM-17-8}}
\hfill{\blue{IFT-UAM/CSIC-17-046}}
\hfill{{\blue{SISSA  24/2017/FISI}}
\hfill\\}
\vskip 1cm
\begin{center}
\mathversion{bold}
{\LARGE\bf Revisiting Minimal Lepton Flavour Violation}\\[4mm]
{\LARGE\bf in the Light of Leptonic CP Violation}\\[4mm]
\mathversion{normal}
\vskip .3cm
\end{center}
\vskip 0.5  cm
\begin{center}
{\large D.N.~Dinh}~$^{a),b)}$,
{\large L.~Merlo}~$^{c)}$,
{\large S.T.~Petcov}~$^{d),e)}$,
{\large R.~Vega-\'Alvarez}~$^{c)}$\\
\vskip .7cm
{\footnotesize
$^{a)}$~
Mathematical and high energy physics group, Institute of physics,\\
Vietnam academy of science and technology,
10 Dao Tan, Ba Dinh, Hanoi, Viet Nam\\
\vskip .1cm
$^{b)}$~Department of Physics, University of Virginia, Charlottesville, VA 22904-4714, USA\\
\vskip .1cm
$^{c)}$~
Departamento de F\'isica Te\'orica and Instituto de F\'isica Te\'orica, IFT-UAM/CSIC,\\
Universidad Aut\'onoma de Madrid, Cantoblanco, 28049, Madrid, Spain\\
\vskip .1cm
$^{d)}$~
SISSA and INFN-Sezione di Trieste, Via Bonomea 265, 34136 Trieste, Italy\\
\vskip .1cm
$^{e)}$~
Kavli IPMU, University of Tokyo (WPI), Tokyo, Japan
\vskip .3cm
\begin{minipage}[l]{.9\textwidth}
\begin{center}
\textit{E-mail:}
\email{dndinh@iop.vast.ac.vn},
\email{luca.merlo@uam.es},
\email{petcov@sissa.it},
\email{roberto.vegaa@estudiante.uam.es}
\end{center}
\end{minipage}
}
\end{center}
\vskip 0.5cm
\begin{abstract}
The Minimal Lepton Flavour Violation (MLFV) framework is discussed after the recent indication for CP violation in the leptonic sector. Among the three distinct versions of MLFV, the one with degenerate right-handed neutrinos will be disfavoured, if this indication is confirmed. The predictions for leptonic radiative rare decays and muon conversion in nuclei are analysed, identifying strategies to disentangle the different MLFV scenarios. The claim that the present anomalies in the semi-leptonic $B$-meson decays can be explained within the MLFV context is critically re-examined concluding that such an explanation is not compatible with the present bounds from purely leptonic processes. 
\end{abstract}
\end{titlepage}
\setcounter{footnote}{0}

\pdfbookmark[1]{Table of Contents}{tableofcontents}
\tableofcontents

\section{Introduction}
\label{Sect:Intro}

The discovery~\cite{Abe:2011sj,Adamson:2011qu,Abe:2011fz,An:2012eh,Ahn:2012nd} of a non-vanishing reactor angle $\theta^\ell_{13}$ in the lepton mixing matrix led to a huge fervour in the flavour community and to a deep
catharsis in the model building approach.

When the value of this angle was still unknown, the closeness to a maximal mixing value of the atmospheric angle $\theta^\ell_{23}$ was suggesting a maximal oscillation between muon- and tau-neutrinos: in terms of symmetries of the Lagrangian acting on the flavour space, it could be described by a discrete Abelian $Z_2$ symmetry, which, in turn, implied a vanishing reactor angle. The simplicity and the elegance of this pattern, i.e. one maximal angle and one vanishing one, convinced part of the community that Nature could have made us a favour and that neutrino physics could indeed be described, at least in the atmospheric and reactor sectors, by this texture~\cite{Fukuyama:1997ky,Altarelli:1998sr}.

 An approach followed for such constructions was to write a Lagrangian whose leading order terms described specific textures for the Yukawa matrices, leading to $\theta^\ell_{13}=0^\circ$ and $\theta^\ell_{23}=45^\circ$. Often, this was done such that the Yukawa matrix for the charged leptons was diagonal while the Yukawa matrix for the light active neutrinos was diagonalised by the so-called Tri-Bimaximal mixing matrix~\cite{Harrison:2002er,Harrison:2002kp,Xing:2002sw}, which  predicts, besides a vanishing reactor mixing angle and a maximal atmospheric one $\theta^\ell_{23}=45^\circ$, a solar angle satisfying to $\sin^2\theta^\ell_{12}=1/3$, in a very good agreement with the neutrino oscillation data.

Pioneer models can be found in Refs.~\cite{Ma:2001dn,Babu:2002dz,Altarelli:2005yp,Altarelli:2005yx,Altarelli:2006kg}, where the discrete non-Abelian group $A_4$ was taken as a flavour symmetry of the lepton sector. Several distinct proposals followed, i) attempting to achieve the Tri-Bimaximal pattern, but with other flavour symmetries (see for example Refs.~\cite{deMedeirosVarzielas:2006fc,Feruglio:2007uu,Bazzocchi:2009pv,Bazzocchi:2009da});
or ii) adopting other mixing patterns to describe neutrino oscillations, such as the Bimaximal mixing\footnote{Bimaximal mixing can be obtained by assuming the existence of an approximate $U(1)$ symmetry corresponding to the conservation of the non-standard lepton charge $L'= L_e - L_{\mu} - L_{\tau}$ and additional discrete $\mu - \tau$ symmetry \cite{Petcov:1982ya}.}~\cite{Vissani:1997pa,Barger:1998ta}, the Golden Ratio mixing~\cite{Kajiyama:2007gx,Rodejohann:2008ir} and the Trimaximal mixing~\cite{King:2011zj}; iii) analysing the possible perturbations or modifications to Bimaximal mixing, Tri-Bimaximal mixing etc., arising from the charged lepton sector \cite{Frampton:2004ud,Romanino:2004ww,Altarelli:2004jb,Hochmuth:2007wq}, vi) implementing the so-called quark-lepton complementarity~\cite{Petcov:1993rk,Minakata:2004xt} which suggests that the lepton and quark sectors should not be treated independently, but a common dynamics could explain both the mixings~\cite{Altarelli:2009gn,Toorop:2010yh,Meloni:2011fx}. Further details could be found for example in these reviews~\cite{Altarelli:2010gt,Grimus:2011fk,Altarelli:2012ss,Bazzocchi:2012st,King:2013eh,King:2015aea}.

 After the discovery of a non-vanishing $\theta^\ell_{13}$ and
the improved sensitivity on the other two mixing angles,
which pointed out that $\theta^\ell_{23}$ best fit is not $45^\circ$ (the most recent global fits on neutrino oscillation data can be found in Refs.~\cite{Capozzi:2016rtj,Esteban:2016qun,Capozzi:2017ipn}), models based on discrete symmetries underwent to a deep rethinking. A few strategies have been suggested: introduction of additional parameters in preexisting minimal models, see for example Refs.~\cite{Ma:2011yi,King:2011ab}; implementation of features that allow sub-leading corrections only in specific directions in the flavour space~\cite{Lin:2009bw,Altarelli:2009kr,Varzielas:2010mp,Altarelli:2012bn}; search for alternative flavour symmetries or mixing patterns that lead already in first approximation to $\theta^\ell_{13}\neq 0^\circ$ and $\theta^\ell_{23}\neq 45^\circ$~\cite{Toorop:2011jn,deAdelhartToorop:2011re}. One can fairly say that the latest neutrino data can still be described in the context of discrete symmetries, but at the prize of
fine-tunings and/or less minimal mechanisms.

Alternative approaches to discrete flavour model building strengthened after 2011 and, in particular, constructions based on continuous symmetries were considered interesting possibilities: models based on the simple $U(1)$ (e.g. Refs.~\cite{Froggatt:1978nt,Altarelli:2000fu,Altarelli:2002sg,Buchmuller:2011tm,Altarelli:2012ia,Bergstrom:2014owa}) or based on $SU(3)$ (e.g. Refs.~\cite{King:2001uz,King:2003rf})  or  the so-called Minimal Flavour Violation (MFV)~\cite{Chivukula:1987py,DAmbrosio:2002vsn}, and its leptonic versions~\cite{Cirigliano:2005ck,Davidson:2006bd,Gavela:2009cd,Alonso:2011jd}, dubbed MLFV. The latter is a setup where the flavour symmetry is identified with the symmetry of the fermionic kinetic terms, or in other words, the symmetry of the SM Lagrangian in the limit of vanishing Yukawa couplings: it is given by products of $U(3)$ factors, one for each fermion spinor of the considered spectrum. Fermion masses and mixings are then described once the symmetry is broken. This approach allows to relate any source of flavour and CP violation in the SM and beyond to the Yukawa couplings, such that any flavour effect can be described in terms of fermion masses and mixing angles. The M(L)FV is not a complete model, as fermion masses and mixings are just described while their origin is not explained (attempts to improve with this respect can be found in Refs.~\cite{Anselm:1996jm,Barbieri:1999km,Berezhiani:2001mh,Feldmann:2009dc,Alonso:2011yg,Nardi:2011st,Alonso:2012fy,Alonso:2013mca,Alonso:2013nca,Fong:2013dnk}). It is instead a framework where observed flavour violating observables are described in agreement with data and unobserved flavour violating signals are not expected to be observed with the current experimental sensitivities, but could be observable in the future planned experiments with significantly higher sensitivity, assuming the New Physics (NP) responsible for these phenomenology at the TeV scale or slightly higher~\cite{DAmbrosio:2002vsn,Cirigliano:2005ck,Cirigliano:2006su,Davidson:2006bd,Grinstein:2006cg,Paradisi:2009ey,Grinstein:2010ve,Feldmann:2010yp,Guadagnoli:2011id,Alonso:2011jd,Buras:2011zb,Buras:2011wi,Alonso:2012jc,Alonso:2012pz,Lopez-Honorez:2013wla,Barbieri:2014tja,Gavela:2009cd,Alonso:2016onw,Crivellin:2016ejn}. \\

The recent indication of a relatively large Dirac CP violation in the lepton sector~\cite{Forero:2014bxa,Blennow:2014sja,Capozzi:2013csa,Capozzi:2016rtj,Esteban:2016qun,Capozzi:2017ipn} represented a new turning point in the sector. Present data prefer a non-zero Dirac CP phase, $\delta^\ell_{CP}$, over CP conservation at more than $2\sigma$'s, depending on the specific neutrino mass ordering. Moreover, the best fit value for the leptonic Jarlskog invariant, $J^\ell_\text{CP}\simeq-0.033$~\cite{Esteban:2016qun}, is numerically much larger in magnitude
than its quark sibling, $J^\ell_\text{CP}\simeq3.04\times 10^{-5}$~\cite{Olive:2016xmw}, indicating potentially a much larger CP violation in the lepton sector than in the quark sector.

In the field of discrete flavour models, this indication translated into looking, for the first time, for approaches and/or contexts where, besides the mixing angles, also the lepton phase(s) were predicted: new models were presented with the CP symmetry as part of the full flavour symmetry~\cite{Feruglio:2012cw,Holthausen:2012dk,Feruglio:2013hia,Girardi:2013sza,Branco:2015gna,Ding:2015rwa,Varzielas:2016zjc}; studies on the mixing patterns and their modifications to provide realistic descriptions of oscillation data were performed~\cite{Shimizu:2014ria,Petcov:2014laa,Girardi:2015vha,Girardi:2015rwa}; an intense activity was dedicated to investigate sum rules involving neutrino masses, mixing angles and $\delta^\ell_{CP}$~\cite{Petcov:2014laa,Girardi:2015vha,Girardi:2015rwa,King:2013psa,Ballett:2013wya,Ballett:2014dua,Girardi:2014faa,Gehrlein:2016wlc}.

The indication for CP violation in the lepton sector also had an impact on models based on continuous flavour symmetries. In particular, one very popular version of MLFV~\cite{Cirigliano:2005ck} strictly requires CP conservation as a working assumption and therefore, if this indication is confirmed, this setup will be disfavoured.

{\it The first goal of this paper is to update previous studies on MLFV in the light of the last global fit on neutrino oscillation data and to discuss the impact of the recent indication for CP violation in the lepton sector.} Indeed, the last studies on MLFV date back to the original papers in 2005~\cite{Cirigliano:2005ck,Davidson:2006bd} and 2011~\cite{Alonso:2011jd}, before the discovery of a non-vanishing $\theta^\ell_{13}$ and lacking any information about the leptonic CP phase.\\

The search for an explanation of the heterogeneity of fermion masses and mixings, the so-called Flavour Puzzle, is just a part of the Flavour Problem of particle physics. A second aspect of this problem is related to the fact that models involving NP typically introduce new sources of flavour violation. Identifying the mechanism which explains why the experimentally measured flavour violation is very much consistent with the SM predictions is a crucial aspect in flavour physics. The use of flavour symmetries turned out to be useful also with this respect: a very well-known example is the MFV setup, as previously discussed, whose construction was originally meant exactly to solve this aspect of the Flavour Problem. Promising results have been obtained also with smaller symmetries than the MFV ones, both continuous~\cite{Barbieri:2011ci,Barbieri:2011fc,Barbieri:2012uh,Barbieri:2012bh,Barbieri:2015yvd,Bordone:2017anc} and discrete~\cite{Feruglio:2008ht,Ishimori:2008uc,Feruglio:2009hu,Feruglio:2009iu,Toorop:2010ex,Toorop:2010kt,Ishimori:2010su,Merlo:2011hw}.

The Flavour Problem becomes even more interesting after the indications for anomalies in the semi-leptonic $B$-meson decays: the angular observable $P'_5$ in the $B\to K^\ast\mu^+\mu^-$ decay presents a tension with the SM prediction of $3.7\sigma$~\cite{Aaij:2013qta,Aaij:2015oid} and $2\sigma$~\cite{Abdesselam:2016llu}, considering LHCb and Belle data, respectively; the Branching Ratio of $B_s\to\phi\mu^+\mu^-$ is in tension with the SM prediction at $3.2\sigma$~\cite{Aaij:2015esa}; the ratio $R_{D^{\ast}}^{\ell}\equiv BR(\ov B\to D^{(\ast)}\tau\ov\nu)_\text{exp}/BR(\ov B\to D^{(\ast)}\ell\ov\nu)_\text{exp}\times BR(\ov B\to D^{(\ast)}\ell\ov\nu)_\text{SM}/ BR(\ov B\to D^{(\ast)}\tau\ov\nu)_\text{SM}$ with $\ell=e,\,\mu$ indicates a $3.9\sigma$ violation of $\tau/\ell$ universality\cite{Fajfer:2012vx,Lees:2013uzd,Na:2015kha,Aaij:2015yra,Huschle:2015rga}; the ratio $R_K\equiv BR(B^+\to K^+\mu^+\mu^-)/BR(B^+\to K^+e^+e^-)$ is in a $2.6\sigma$ tension with the SM prediction~\cite{Aaij:2014ora}, indicating lepton universality violation in the $e/\mu$ sector. The latter has been confirmed also by the recent announcement of the measure of $R_{K^\ast}\equiv BR(B^0\to K^{\ast0}\mu^+\mu^-)/BR(B^0\to K^{\ast0}e^+e^-)$ is in a $2.4-2.5\sigma$ ($2.2-2.4\sigma$) tension with the SM prediction in the central-$q^2$ region (low-$q^2$ region)\cite{BifaniTalk}. Under the assumption that these anomalies are due to NP, and not due to an underestimation of the hadronic effects~\cite{Lyon:2014hpa,Descotes-Genon:2014uoa,Jager:2014rwa,Ciuchini:2015qxb,Capdevila:2017ert,Chobanova:2017ghn} or due to a statistical fluctuation, a global analysis on $b\to s$ data can attempt to identify the properties of the underlying theory. Adopting an effective description, these results can be translated into constraints of the Wilson coefficients of the Hamiltonian describing $\Delta B=1$ decays: the results of such analysis~\cite{Descotes-Genon:2013wba,Altmannshofer:2013foa,Hurth:2013ssa,Ghosh:2014awa,Altmannshofer:2014rta,Descotes-Genon:2015uva,Hurth:2016fbr,Capdevila:2016ivx,Altmannshofer:2017fio,Capdevila:2017bsm,Altmannshofer:2017yso,Geng:2017svp,Ciuchini:2017mik} are that the anomalies can be explained with a modification of the Wilson coefficients $C_9$ and $C_{10}$ defined as
\be
\HH^\text{eff}_{\Delta B=1}\supset -\dfrac{4G_F}{\sqrt2}\dfrac{e^2}{(4\pi)^2}V_{tb}V^\ast_{ts}\,\Big[\ov s \gamma_\mu P_L b\Big]\Big[\ov\ell\gamma^\mu\left(C_9+C_{10}\gamma_5\right)\ell\Big]+\hc
\label{btosHamiltonian}
\ee
where $V$ is the CKM matrix, $P_L=(1-\gamma_5)/2$ is the usual left-handed (LH) chirality projector, $b$ and $s$ refer to the bottom and strange quarks, respectively, $\ell$ are the charged leptons, and the pre-factors refer to the traditional normalisation. Writing each of the coefficients as the sum of the purely SM contribution and the NP one, $C_i=C^\text{SM}_i+\delta C_i$, the results of a one-operator-at-a-time analysis~\cite{Altmannshofer:2017yso} suggest lepton universality violation in the $e/\mu$ sector quantifiable in
\be
\delta C^e_9=-\delta C^e_{10}\in[+0.56,\,+1.02]\quad \text{and} \quad \delta C^\mu_9=-\delta C^\mu_{10}\in[-0.81,\,-0.48]\quad @1\sigma\,,
\label{ValuesC9C10}
\ee
corresponding to $4.3\sigma$ and $4.2\sigma$ tension with the SM predictions, respectively.

The hypothetical underlying theory, which manifests itself at low energies with these features, will necessarily respect the SM gauge invariance, and therefore will also contribute to $b\to c$ processes and hopefully solve the $R^\ell_{D^{(\ast)}}$ anomalies.

Several attempts have been presented in the literature to explain the deficit on $C_9$ and/or $C_{10}$, including the MLFV approach: Ref.~\cite{Lee:2015qra} considers the version of MLFV introduced in Ref.~\cite{Cirigliano:2005ck} and constraints on the Lagrangian parameters and on the Lepton Flavour Violating (LFV) scale have been obtained requiring to reproduce the values of $\delta C^e_9$ and $\delta C^e_{10}$ aforementioned.

{\it A second goal of this paper is to revisit the results presented in Ref.~\cite{Lee:2015qra} considering the constraints from purely leptonic observables, such as radiative rare decays and $\mu\to e$ conversion in nuclei. Moreover, the analysis will be extended to the other versions of MLFV~\cite{Alonso:2011jd}.}

The structure of the paper can easily be deduced from the table of content: first, in Sect.~\ref{Sect:MFV}, basic concepts of MFV and MLFV will be recalled, underlying the differences between the distinct versions of MLFV; then, in Sect.~\ref{Sect:LeptoPhem}, several processes in the lepton sector will be discussed considering the last global fit on neutrino data and the recent indication for leptonic CP violation; in Sect.~\ref{Sect:bAnomalies}, the anomalies in the $b\to s$ decays will be discussed, pointing out the differences with respect to previous literature; finally, concluding remarks will be presented in Sect.~\ref{Sect:Conc}.

\section{Minimal (Lepton) Flavour Violation}
\label{Sect:MFV}

If a theory of NP, with a characteristic scale of a few TeVs, behaves at low energy accordingly to the MFV ansatz, i.e. the SM Yukawa couplings are the only sources of flavour and CP violation even beyond the SM, then its flavour protection is guaranteed: the large majority of observed flavour processes in the quark sector are predicted in agreement with data~\cite{DAmbrosio:2002vsn,Cirigliano:2006su,Hurth:2008jc,Paradisi:2009ey,Lalak:2010bk,Grinstein:2010ve,Feldmann:2010yp,Guadagnoli:2011id,Redi:2011zi,Buras:2011zb,Buras:2011wi,Alonso:2012jc,Hurth:2012jn,Alonso:2012pz,Lopez-Honorez:2013wla,Calibbi:2013mka,Bishara:2015mha}; unseen flavour changing processes, for example leptonic radiative rare decays, are predicted to have strengths which are inside the present experimental sensitivity~\cite{Cirigliano:2005ck,Davidson:2006bd,Alonso:2011jd,Redi:2013pga,He:2014efa,Feruglio:2015gka,Feldmann:2016hvo,Alonso:2016onw}.

In the modern realisation of the MFV ansatz, the flavour symmetry corresponds to the one arising in the limit of vanishing Yukawa couplings. This massless Lagrangian is left invariant under a tridimensional unitary transformations in the flavour space associated to each fermion spinor. In the quark sector, it is given by
\be
\G_Q\times U(1)_B\times U(1)_{A^u}\times U(1)_{A^d}
\qquad\text{with}\qquad
\G_Q= SU(3)_{q_L}\times SU(3)_{u_R}\times SU(3)_{d_R}\,,
\label{GFQ}
\ee
where $q_L$ refer to the $SU(2)_L$-doublet of quarks, and $u_R$ and $d_R$ to the $SU(2)_L$-singlets. The Abelian terms can be identified with the Baryon number, and with two axial rotations, in the up- and down-quark sectors respectively, which do not distinguish among the distinct families~\cite{RodrigosThesis}. On the contrary, the non-Abelian factors rule the interactions among the generations and govern the amount of flavour violation: they are the key ingredients
of MFV and will be in the focus of the analysis in which follows.

The explicit quark transformations read
\be
\begin{array}{ccc}
q_L\sim ({\bf 3},\,1,\,1)_{\G_Q} \qquad\qquad\qquad
&u_R\sim (1,\,{\bf 3},\,1)_{\G_Q}\qquad \qquad\qquad
&d_R\sim (1,\,1,\,{\bf 3})_{\G_Q} \\[2mm]
q_L\to \U_{q_L}q_L \qquad\qquad\qquad
&u_R\to\U_{u_R}u_R\qquad \qquad\qquad
&d_R\to\U_{d_R}d_R\,, \\
\end{array}
\ee
where $\U_i\in SU(3)_i$ are $3\times 3$ unitary matrices acting in the flavour space. The quark Lagrangian is invariant under these transformations, except for the Yukawa interactions:
\be
\LL_Q=-\ov q_L Y_u \tilde H u_R-\ov q_L Y_d H d_R + \hc\,,
\ee
where $Y_i$ are $3\times 3$ matrices in the flavour space, $H$ is the $SU(2)_L$-double Higgs field, and $\tilde H=i\sigma_2 H^\ast$. $\LL_Q$ can be made invariant under $\G_Q$ promoting the Yukawa matrices to be spurion fields,  i.e. auxiliary non-dynamical fields, denoted by $\Y_u$ and $\Y_d$, with specific transformation properties under the flavour symmetry:
\be
\begin{array}{cc}
\Y_u\sim({\bf 3}, \ov {\bf 3}, 1)_{\G_Q}\qquad\qquad\qquad
&\Y_d\sim({\bf 3}, 1, \ov {\bf 3})_{\G_Q}\\[2mm]
\Y_u\to \U_{q_L}\,\Y_u\, \U^\dag_{u_R}\qquad\qquad\qquad
&\Y_d\to \U_{q_L}\,\Y_d\, \U^\dag_{d_R}\,.
\end{array}
\ee
Once the Yukawa spurions acquire a background value, the flavour symmetry is broken and in consequence fermions masses and mixings are generated. A useful choice for these background values is to identify them with the SM Yukawa couplings: in a given basis, $Y_d$ is diagonal and describes only down-type quark masses, while $Y_u$ contains non-diagonal entries and accounts for both up-type quark masses and the CKM matrix $V$:
\be
\begin{gathered}
\vev{\Y_u}\equiv Y_u=\dfrac{\sqrt2}{v}V^\dag\hat M_u
\,,\qquad\qquad
\vev{\Y_d}\equiv Y_d=\dfrac{\sqrt2}{v}\hat M_d
\,,
\end{gathered}
\label{YukawaValues}
\ee
where $v=246$ GeV is the Higgs vacuum expectation value (VEV) defined by $\vev{H^0}=v/\sqrt2$, and $\hat M_{u,d}$ are the diagonal mass matrices for up- and down-type quarks,
\be
\hat M_u\equiv\diag(m_u,\,m_c,\,m_t)\,,\qquad\qquad
\hat M_d\equiv\diag(m_d,\,m_s,\,m_b)\,.
\ee

When considering low-energy flavour processes, they can be described within the effective field theory approach through non-renormalisable operators suppressed by suitable powers of the scale associated to the messenger of the interaction. These structures could violate the flavour symmetry $\G_Q$, especially if they describe flavour changing observables. As for the Yukawa Lagrangian, a technical way out to recover flavour invariance is to insert powers of the Yukawa spurions. Once the spurions acquire background values, the corresponding processes are predicted in terms of quark masses and mixings. Several studies already appeared addressing this topic~\cite{DAmbrosio:2002vsn,Cirigliano:2006su,Hurth:2008jc,Paradisi:2009ey,Lalak:2010bk,Grinstein:2010ve,Feldmann:2010yp,Guadagnoli:2011id,Redi:2011zi,Buras:2011zb,Buras:2011wi,Alonso:2012jc,Hurth:2012jn,Alonso:2012pz,Lopez-Honorez:2013wla,Calibbi:2013mka,Bishara:2015mha} and, as already mentioned at the beginning of this section, the
  results
  show that flavour data in the quark sector are well described within the MFV(-like) approach. Indeed, the Yukawa spurions act as expanding parameters and processes described by effective operators with more insertions of the spurions obtain stronger suppressions\footnote{The top Yukawa represents an exception as it cannot be technically taken as an expanding parameter. This aspect has been treated in Refs.~\cite{Kagan:2009bn}, where a resummation procedure has been illustrated.}.

MFV, however, cannot be considered a complete flavour model, as there is not explanation of the origin of quark masses and mixings. There have been attempts to go from the effective-spurionic approach to a more fundamental description, promoting the Yukawa spurions to be dynamical fields, called flavons, acquiring a non-trivial VEV. The corresponding scalar potentials have been discussed extensively with interesting consequences~\cite{Anselm:1996jm,Barbieri:1999km,Berezhiani:2001mh,Feldmann:2009dc,Alonso:2011yg,Nardi:2011st,Fong:2013dnk}: a conclusive dynamical justification for quark masses and mixing is still lacking, but the results are encouraging as the potential minima lead, at leading order, to non-vanishing masses for top and bottom quarks and to no mixing.

\subsection{The Lepton Sector}

The lepton sector is more involved with respect to the quark one, due to the lack of knowledge on neutrino masses: indeed, while the charged lepton description mimics the one of down-quarks, light active neutrino masses, and then the lepton mixing, cannot be described within the SM.

Several ways out have been presented in the literature to provide a description for the lepton sector, and the focus here will be on two well-defined approaches, one maintaining the SM spectrum but relaxing the renormalisability criterium, and the other adding new particles in a still renormalisable theory.

\subsubsection*{Minimal Field Content (MFC)}
Giving up with renormalisability, active neutrino masses can be described via the so-called Weinberg operator~\cite{Weinberg:1979sa}, a non-renormalisable operator of canonical dimension 5 which breaks explicitly Lepton number by two units,
\be
\O_W=\dfrac{1}{2}\left(\ov{\ell^c_L}\tilde{H}^\ast\right)\dfrac{g_\nu}{\LLN}\left(\tilde H^\dag\ell_L\right)\,,
\label{WeinbergOp}
\ee
where $\ell_L^c\equiv C \ov{\ell_L}^T$, $C$ being the charge conjugation matrix
($C^{-1} \gamma_{\mu} C = -\, \gamma^T_{\mu}$),
$g_\nu$ is an adimensional symmetric $3\times3$ matrix in the flavour space
and $\LLN$ is the scale of Lepton Number Violation (LNV).
The flavour symmetry arising from the kinetic terms in this case is
\be
\G_L\times U(1)_L\times U(1)_{A^e}\qquad\qquad
\text{with}\qquad\qquad
\G_L=SU(3)_{\ell_L}\times SU(3)_{e_R}\,,
\label{FullGL}
\ee
where $U(1)_L$ is the Lepton number while $U(1)_{A^e}$ is an axial rotation in $\ell_L$ and $e_R$, and the non-Abelian transformations of the leptons read
\be
\begin{array}{cc}
\ell_L\sim ({\bf 3},\,1)_{\G_L} \qquad\qquad\qquad
&e_R\sim (1,\,{\bf 3})_{\G_L} \\[2mm]
\ell_L\to \U_{\ell_L}\ell_L \qquad\qquad\qquad
&e_R\to\U_{e_R}e_R\,.
\end{array}
\ee
The part of the Lagrangian describing lepton masses and mixings,
\be
\LL_L= -\ov \ell_L Y_e H e_R - \O_W + \hc\,,
\ee
is not invariant under $\G_L$, but this can be cured by promoting $Y_e$ and $g_\nu$ to be spurion fields, $\Y_e$ and $\g_\nu$, transforming as
\be
\begin{array}{cc}
\Y_e\sim({\bf 3},\,\ov{\bf 3})_{\G_L}\qquad\qquad\qquad
&\g_\nu \sim (\ov{\bf 6},1)_{\G_L}\\[2mm]
\Y_e\to \U_{\ell_L}\,\Y_e\, \U^\dag_{e_R}\qquad\qquad\qquad
&\g_\nu\to \U^\ast_{\ell_L}\,\g_\nu\, \U^\dag_{\ell_L}\,.
\end{array}
\ee
Lepton masses and the PMNS matrix $U$ arise once $\Y_e$ and $\g_\nu$ acquire a background value that can be chosen to be
\be
\vev{\Y_e}\equiv Y_e=\dfrac{\sqrt2}{v}\hat M_\ell\,,\qquad\qquad
\vev{\,\g_\nu}\equiv g_\nu =\dfrac{2\LLN}{v^2}U^\ast\hat M_\nu
U^\dag\,,
\label{BackgroudSpurionsMFC}
\ee
with $\hat M_{\ell,\nu}$ being the diagonal matrices
of the charged lepton and active neutrino mass eigenvalues,
\be
\hat M_\ell\equiv \diag(m_e,\,m_\mu,\,m_\tau)\,,\qquad\qquad
\hat M_\nu\equiv \diag\left(m_{\nu_1},\,m_{\nu_2},\,m_{\nu_3}\right)\,,
\ee
and $U$ defined as the product of four matrices~\cite{Olive:2016xmw},
\be
U=R_{23}(\theta^\ell_{23})\cdot R_{13}(\theta^\ell_{13},\delta^\ell_{CP})\cdot R_{12}(\theta^\ell_{12})\cdot \diag\left(1,e^{i\frac{\alpha_{21}}{2}},e^{i\frac{\alpha_{31}}{2}}\right)\,,
\ee
with $R_{ij}(\theta^\ell_{ij})$ a generic rotation of the angle $\theta^\ell_{ij}$ in the $ij$ sector, with the addition of the Dirac CP phase $\delta^\ell_{CP}$ in the reactor sector, and $\alpha_{21,31}$ the Majorana phases 
\cite{Bilenky:1980cx}.

As discussed for the quark case, $Y_e$ and $g_\nu$ act as expanding parameters: operators with more insertions of these spurions describe processes that receive stronger suppressions. This perturbative treatment requires, however, that the largest entries in $Y_e$ and $g_\nu$ are at most $\O(1)$. The charged lepton Yukawa satisfies to this condition as the largest entry is $\sim m_\tau/v$. The neutrino spurion $g_\nu$ is instead function of $\LLN$:
requiring that $|g_{\nu\,ij}|<1$ leads to an upper bound on the LNV scale,
which depends on $|(U^\ast\hat M_\nu U^\dag)_{ij}|$ that is a function of the type of neutrino mass spectrum (NO or IO), of the value of the lightest neutrino mass and of the values of the Majorana and Dirac CP violation phases.
The lowest upper bound is given approximately by:
\be
\LLN\simeq\dfrac{v^2}{2} \dfrac{g_\nu}{\sqrt{\Dmatm}}\lesssim 6\times 10^{14}\GeV\,.
\label{LNVScaleBoundMFC}
\ee

It will be useful for the phenomenological discussion in the next sections to remember that the spurion combination $\g_\nu^\dag\, \g_\nu$ transforms as $({\bf 8},\,1)_{\G_L}$ and to introduce the quantity
\be
\Delta\equiv g_\nu^\dag g_\nu=\dfrac{4\Lambda^2_L}{v^4}U\hat M^2_\nu
U^\dag\,.
\label{DeltaMFC}
\ee

\subsubsection*{Extended Field Content (EFC)}
Enlarging the SM spectrum by the addition of three RH neutrinos $N_R$ leads to the so-called type I Seesaw context~\cite{Minkowski:1977sc,GellMann:1980vs,Yanagida:1980xy,Mohapatra:1980yp,Schechter:1980gr}, described by the following Lagrangian:
\be
\LL_\text{L--SS}=-\ov \ell_L Y_e H e_R-\ov \ell_L Y_\nu \tilde H N_R-\dfrac{1}{2}\muLN\ov{N^c_R}Y_N N_R+\hc\,,
\label{LagSS}
\ee
where $Y_e$, $Y_\nu$ and $Y_N$ are adimensional $3\times3$ matrices in the flavour space, while $\muLN$ stands for the scale of Lepton number violation, broken by two units by the last term on the right of this equation. Assuming a hierarchy between $\muLN$ and $v$, $\muLN\gg v$, it is then possible to easily block-diagonalise the full $6\times 6$ neutrino mass matrix, and obtain the induced masses for the light active neutrinos: in terms of the parameter $g_\nu$ appearing in the Weinberg operator in Eq.~(\ref{WeinbergOp}), they are given by
\be
\dfrac{g^\dag_\nu}{\LLN}=Y_\nu\dfrac{Y_N^{-1}}{\muLN}Y_\nu^T\,.
\label{g_nu}
\ee

The fermionic kinetic terms of the SM extended with 3 RH neutrinos manifest the following flavour symmetry:
\be
\G_L\times U(1)_L\times U(1)_{A^e}\times U(1)_{A^N}\quad
\text{with}\quad
\G_L=SU(3)_{\ell_L}\times SU(3)_{e_R}\times SU(3)_{N_R}\,,
\label{FullGLSS}
\ee
under which leptons transform as
\be
\begin{array}{ccc}
\ell_L\sim ({\bf 3},\,1,\,1)_{\G_L} \qquad\qquad\qquad
&e_R\sim (1,\,{\bf 3},\,1)_{\G_L} \qquad\qquad\qquad
&N_R\sim (1,\,1,\,{\bf 3})_{\G_L} \\[2mm]
\ell_L\to \U_{\ell_L}\ell_L \qquad\qquad\qquad
&e_R\to\U_{e_R}e_R \qquad\qquad\qquad
&N_R\to\U_{N_R}N_R\,,
\end{array}
\label{NewEqEFCI}
\ee
and where $U(1)_{A^N}$ is an axial transformation associated to $N_R$ and $SU(3)_{N_R}$ is a new rotation that mixes the  three RH neutrinos. The Lagrangian in Eq.~(\ref{LagSS}) breaks explicitly $\G_L$ defined in Eq.~(\ref{FullGLSS}), but the invariance can be technically restored promoting $Y_E$, $Y_\nu$ and $Y_N$ to be spurions fields, $\Y_E$, $\Y_\nu$ and $\Y_N$, transforming as
\be
\begin{array}{ccc}
\Y_e\sim({\bf 3},\,\ov{\bf 3},\,1)_{\G_L}\qquad\qquad
&\Y_\nu\sim({\bf 3},\,1,\,\ov{\bf 3})_{\G_L}\qquad\qquad
&\Y_N \sim (1,\,1,\ov{\bf 6})_{\G_L}\\[2mm]
\Y_e\to \U_{\ell_L}\,\Y_e\, \U^\dag_{e_R}\qquad\qquad
&\Y_\nu\to \U_{\ell_L}\,\Y_\nu\, \U^\dag_{N_R}\qquad\qquad
&\Y_N\to \U^\ast_{N_R}\,\Y_N\, \U^\dag_{N_R}\,.
\end{array}
\ee
Lepton masses and mixing are then described when these spurion fields acquire the following background values:
\be
\vev{\Y_e}\equiv Y_e=\dfrac{\sqrt2}{v}\hat M_\ell\,,\qquad\qquad
\vev{\Y_\nu}\vev{\Y_N^{-1}}\vev{\Y_\nu^T}\equiv Y_\nu Y_N^{-1}Y_\nu^T=\dfrac{2\muLN}{v^2}U\hat M_\nu U^T\,.
\label{VEVsSS}
\ee
Differently from the quark sector and the MFC lepton case, it is not possible to identify a unique choice for
$\vev{\Y_\nu}$ and $\vev{\Y_N}$, as only the specific combination in Eq.~(\ref{VEVsSS}) can be associated to the neutrino mass eigenvalues and the PMNS matrix entries. This is a relevant aspect as it nullifies the MLFV flavour protection. Indeed, the basic building blocks for several processes, such as radiative leptonic decays or leptonic conversions, are fermionic bilinears of the type $\ov{\ell}_L^i\Gamma \ell_L^j$, $\ov{\ell}_L^i\Gamma \ell_L^{c\,j}$, $\ov{\ell}_L^i\Gamma e_R^j$ and $\ov{e}_R^i\Gamma e_R^j$, with $\Gamma$ standing for combination of Dirac $\gamma$ matrices and/or Pauli $\sigma$ matrices. In the unbroken phase, these terms are invariant under the flavour symmetry contracting the flavour indices with combinations of the spurions transforming as $({\bf 8},\,1,\,1)_{\G_L}$, $({\bf 6},\,1,\,1)_{\G_L}$, $({\bf 3},\,\ov{\bf 3},\,1)_{\G_L}$, and $(1,\,{\bf 8},\,1)_{\G_L}$, among others. These spurion combinations are distinct from the combination of $\Y_\nu$ and 
 $\Y_N$ t
 hat appears in Eq.~(\ref{VEVsSS}): a few examples are
\be
\begin{aligned}
({\bf 8},\,1,\,1)_{\G_L}\qquad\qquad
&\Y_\nu \Y_\nu^\dag \,,\,\, \Y_e \Y_e^\dag \,,\,\, \Y_\nu \Y_N^\dag \Y_N \Y_\nu^\dag,\,\,\left(\Y_\nu \Y_\nu^\dag\right)^2,\,\,\ldots\\
({\bf 6},\,1,\,1)_{\G_L}\qquad\qquad
&\Y_\nu \Y_N^\dag \Y_\nu^T \,,\,\, \Y_\nu \Y_N^\dag\Y_N\Y_N^\dag \Y_\nu^T,\,\,\Y_\nu \Y_N^\dag \Y_\nu^T \Y_\nu^\ast \Y_\nu^T\,,\,\,\ldots
\\
({\bf 3},\,\ov{\bf 3},\,1)_{\G_L}\qquad\qquad
&\Y_e,\,\, \Y_\nu\Y_\nu^\dag\Y_e \,,\,\, \Y_e\Y_e^\dag\Y_e,\,\, \Y_\nu \Y_N^\dag \Y_N \Y_\nu^\dag\Y_e,\,\,\ldots\\
(1,\,{\bf 8},\,1)_{\G_L}\qquad\qquad
&\Y_e^\dag \Y_e,\,\, \Y_e^\dag\Y_\nu\Y_\nu^\dag \Y_e,\,\,\Y_e^\dag\Y_\nu\Y_N^\dag \Y_N\Y_\nu^\dag \Y_e,\,\,\ldots
\end{aligned}
\label{FCCombinations}
\ee
In consequence, one concludes that it is not possible to express any flavour changing process involving leptons in terms of lepton masses and mixings, losing in this way the predictive power of MLFV.

This problem can be solved, and predictivity can be recovered, if all the information of neutrino masses and mixing would be encoded into only one spurion background among $Y_\nu$ and $Y_N$, being the other proportional to the identity matrix. Technically, this corresponds to break $\G_L$ following two natural criteria.
\begin{itemize}
\item[I):] $\G_L\to SU(3)_{\ell_L}\times SU(3)_{e_R}\times SO(3)_{N_R}\times CP$~\cite{Cirigliano:2005ck,Davidson:2006bd}.\\[5pt]
Under the assumption that the three RH neutrinos are degenerate in mass, i.e. \mbox{$Y_N\propto \unity$}, $SO(3)_{N_R}$ is broken down to $SO(3)_{N_R}$ and the transformation $\U_{N_R}$ in Eq.~(\ref{NewEqEFCI}) is then an orthogonal matrix. The additional assumption of no CP violation in the lepton sector  is meant to force $Y_e$ and $Y_\nu$ to be real\footnote{Strictly speaking, the condition of CP conservation in the leptonic sector forces the Dirac CP phase to be equal to $\delta^\ell_{CP}=\{0,\,\pi\}$ and the Majorana CP phases to be $\alpha_{21,31}=\{0,\,\pi,\,2\pi\}$. However, $Y_\nu$ is real only if $\alpha_{21,31}=\{0,\,2\pi\}$, and therefore $\alpha_{21,31}=\pi$ needs to be disregarded in order to guarantee predictivity. The CP conservation condition assumed in this context is then stronger than the strict definition.}. With this simplifications, all flavour changing effects involving leptons can be written in terms of $Y_\nu Y_\nu^T$ and $Y_e$, as can be easily deduced from Eq.~(\ref{FCCombinations}). In this case, Eq.~(\ref{VEVsSS}) simplifies to
\be
Y_\nu Y_\nu^T=\dfrac{2\muLN}{v^2}U\hat M_\nu U^T\equiv \Delta\,,
\label{DeltaEFCI}
\ee
eventually redefining $\muLN$ by reabsorbing the norm of $Y_N$, and therefore any flavour changing process can be described in terms of lepton masses and mixings. The last equivalence in the previous equation is a definition that will be useful in the phenomenological analysis.

As for the MFC case, requiring that the spurions respect the perturbativity regime leads to an upper bound on the LNV scale:
\be
\muLN\simeq\dfrac{v^2}{2} \dfrac{Y_\nu Y_\nu^T}{\sqrt{\Dmatm}}\lesssim 6\times 10^{14}\GeV\,,
\label{LNVScaleBoundEFCI}
\ee
numerically the same as the one in Eq.~(\ref{LNVScaleBoundMFC}).

\item[II):] $\G_L\to SU(3)_{\ell_L+N_R}\times SU(3)_{e_R}$~\cite{Alonso:2011jd}.\\[5pt]
Assuming that the three RH neutrinos transform as a triplet under the same symmetry group of the lepton doublets, 
\be
\ell_L, N_R\sim ({\bf 3},\,1)_{\G_L} \qquad\qquad
e_R\sim (1,\,{\bf 3})_{\G_L}\,,
\ee
then the Schur's Lemma guarantees that $\Y_\nu$ transforms as a singlet of the symmetry group and then $Y_\nu$ is a unitary matrix~\cite{Bertuzzo:2009im,AristizabalSierra:2009ex}, which can always be rotated to the identity matrix by a suitable unitary transformation acting only on the RH neutrinos. The only sensible quantities in this context are $\Y_e$ and $\Y_N$, which now transform as 
\be
\Y_e\sim({\bf 3},\,\ov{\bf 3})_{\G_L}\qquad\qquad
\Y_N \sim (\ov{\bf 6},\,1)_{\G_L}\,.
\ee
The background value of $\Y_N$ would eventually encode the norm of $Y_\nu$, in order to consistently take $Y_\nu=\unity$. In this basis, neutrino masses and the lepton mixing are encoded uniquely into $Y_N$,
\be
Y_N=\dfrac{v^2}{2\muLN}U^\ast\hat M_\nu^{-1}U^\dag\,.
\ee
Moreover, all the spurion combinations in Eq.~(\ref{FCCombinations}) can be written only in terms of $Y_e$ and $Y_N$ 
and therefore any flavour changing process can be predicted in terms of lepton masses and mixing. It will be useful in the phenomenological analysis that follows to introduce the quantity
\be
\Delta\equiv Y_N^\dag Y_N =\dfrac{v^4}{4\mu^2_L}U\hat M_\nu^{-2}U^\dag\,.
\label{DeltaEFCII}
\ee

Contrary to what occurs in the MFC and the EFCI cases, the perturbativity condition on $Y_N$ allows to extract a lower bound on the LNV scale:
\be
\muLN\simeq \dfrac{v^2}{2}\dfrac{Y_N^{-1}}{\sqrt{\Dmatm}}\gtrsim 6\times 10^{14}\GeV\,.
\label{LNVScaleBoundEFCII}
\ee
\end{itemize}

Similarly to what discussed for the quark sector, none of the two versions of the MLFV provide an explanation for the origin of lepton masses and mixing, and therefore cannot be considered complete models. In Refs.~\cite{Alonso:2012fy,Alonso:2013mca,Alonso:2013nca} attempts have been presented to provide a dynamical explanation for the flavour puzzle in the lepton sector: as for
the quark sector, the results are not conclusive,
but highlighted interesting features. Indeed, for the MLFV version
with an $SO(3)_{N_R}$ symmetry factor associated to the RH neutrinos,
the minima of the scalar potential, constructed by promoting
$\Y_e$ and $\Y_\nu$ to be dynamical fields, allow a maximal mixing and
a relative maximal Majorana CP phase between two almost degenerate
neutrino mass eigenvalues. This seems to suggest that the large angles
in the lepton sector could be due to the Majorana nature of neutrinos,
in contrast with the quark sector where this does not occur.

No dedicated analysis of the scalar potential arising in the second version of MLFV has appeared in the literature, although the results are not expected to be much different from the ones in the quark sector. However, as a conclusive mechanism to explain lepton masses and mixing is still lacking, both the versions of MLFV remain valid possibilities.\\

As anticipated in Sect.~\ref{Sect:Intro}, the recent indication for a relatively large leptonic CP violation, if confirmed, would disfavour EFCI, due to the required reality of $Y_\nu$. However, in the present discussion and in the analysis that follows, EFCI will not be discarded yet, as the assumption of CP conservation is a distinctive feature of this low-energy description of the lepton sector, but could be avoided in more fundamental ones. Indeed, a model constructed upon the {\it gauged} lepton flavour symmetry $SU(3)_{\ell_L}\times SU(3)_{e_R}\times SO(3)_{N_R}$, without any further hypothesis on CP in the lepton sector, is shown in Ref.~\cite{Alonso:2016onw} to be as predictive as EFCI: indeed, with the Dirac CP phase taken at its best fit value, this gauged flavour model presents several phenomenological results similar to the ones  of EFCI discussed in Ref.~\cite{Cirigliano:2005ck,Davidson:2006bd}. This motivates to consider EFCI as a valid context to describe lepton flavour observables, even if results which show a strong dependence on the value of the Dirac CP phase should be taken with a grain of salt.

\section{Phenomenology in the Lepton Sector }
\label{Sect:LeptoPhem}

In this section, the phenomenology associated to the MFC, EFCI and EFCII cases will be discussed considering specifically leptonic radiative rare decays and $\mu\to e$ conversion in nuclei. While these analyses have already been presented in the original MLFV papers~\cite{Cirigliano:2005ck,Davidson:2006bd,Alonso:2011jd}, in the review part of the present paper the latest discovered value of the reactor angle and the recent indication of non-vanishing CP phase in the leptonic sector will be considered.

The input data that will be used in what follows are the PDG values for the charged lepton masses~\cite{Olive:2016xmw}
\be
m_e=0.51\text{\! MeV}\,,\qquad\qquad
m_\mu=105.66\text{\! MeV}\,,\qquad\qquad
m_\tau=1776.86\pm0.12\text{\! MeV}\,,
\ee
where the electron and muon masses are taken without errors as the sensitivities are negligible, and the results of the neutrino oscillation fit from Ref.~\cite{Esteban:2016qun} reported in Table~\ref{TableOscFit}.

\begin{table}[htp]
\begin{center}
\begin{tabular}{l|c|c|}
										& Normal Ordering 				& Inverted Ordering \\[1mm]
\hline
&&\\
$\sin^2\theta^\ell_{12}$						& $0.306\pm0.012$				& $0.306\pm 0.012$ \\[3mm]
$\sin^2\theta^\ell_{23}$						& $0.441^{+0.027}_{-0.021}$		& $0.587^{+0.020}_{-0.024}$ \\[3mm]
$\sin^2\theta^\ell_{13}$						& $0.02166\pm0.00075$			& $0.02179\pm0.00076$\\[3mm]
$\delta^\ell_{CP}/^\circ$							& $261^{+51}_{-59}$				& $277^{+40}_{-46}$\\[3mm]
$\Delta m^2_{sol}/10^{-5}\text{\! eV}^2$		& $7.50^{+0.19}_{-0.17}$			& $7.50^{+0.19}_{-0.17}$\\[3mm]
$\Delta m^2_{atm}/10^{-3}\text{\! eV}^2$		& $2.524^{+0.039}_{-0.040}$		& $2.514^{+0.0.38}_{-0.041}$\\[3mm]
\hline
\end{tabular}
\end{center}
\caption{\footnotesize\it Three-flavour oscillation parameters from the global fit in Ref.~\cite{Esteban:2016qun}. The results in the second and third columns refer to the Normal and the Inverted Orderings, respectively. The notation has been chosen such that $\Dmsol\equiv m^2_{\nu_2}- m^2_{\nu_1}$, and $\Dmatm\equiv m^2_{\nu_3}-m^2_{\nu_1}$ for NO and $\Dmatm\equiv m^2_{\nu_2}-m^2_{\nu_3}$ for IO. The errors reported correspond to the $1\sigma$ uncertainties.}
\label{TableOscFit}
\end{table}%

The value of the lightest neutrino mass and the neutrino mass ordering are still unknown. For this reason, the results of this section will be discussed in terms of the values of the lightest neutrino mass and for both the Normal Ordering (NO) and the Inverted Ordering (IO). The measured parameters are taken considering their $2\sigma$ error bands~\footnote{EW running effects\cite{Antusch:2003kp,Antusch:2005gp,Ellis:2005dr,Lin:2009sq} are negligible in the analysis presented here.}: this is to underly the impact of the raising indication for a leptonic CP violation.

\subsection{The LFV Effective Lagrangian}

The rates of charged LFV processes, i.e. $\mu\to e+\gamma$, $\mu \to 3e$,  and $\mu\to e$ conversion in nuclei among others, are predicted to be unobservably small in the minimal extension of the SM with light massive Dirac neutrinos, in which the total lepton charge is conserved \cite{Petcov:1976ff}. As a consequence, the rates of such processes have a remarkable sensitivity to NP contributions.

The main observables that will be discussed here are lepton radiative rare decays and $\mu\to e$ conversion in nuclei. Other leptonic observables which are typically very sensible to NP are $\ell\to\ell'\ell'\ell''$ decays, and especially the $\mu\to3 e$ decay, given the significant increase of the sensitivity of the planned experiments. However, these processes do not provide additional information for the results that will be obtained in the following, and therefore they will not be further considered.

Assuming the presence of new physics at the scale $\LLF$ responsible for these observables characterised by a much lower typical energy, one can adopt the description in terms of an effective Lagrangian\footnote{The effective Lagrangian reported here corresponds to the linearly realised EWSB. An alternative would be to considered a non-linear realisation and the corresponding effective Lagrangian dubbed HEFT~\cite{Feruglio:1992wf,Contino:2010mh,Alonso:2012px,Buchalla:2013rka,Brivio:2016fzo,deFlorian:2016spz}. In this context, however, a much larger number of operators should be taken into consideration and a slightly different phenomenology is expected~\cite{Brivio:2013pma,Brivio:2014pfa,Gavela:2014vra,Alonso:2014wta,Hierro:2015nna,Brivio:2015kia,Gavela:2016bzc,Merlo:2016prs,Brivio:2017ije,Hernandez-Leon:2017kea}. The focus in this paper is on the linear EWSB realisation and therefore the HEFT Lagrangian will not be considered in what follows.}: the relevant terms are then given by\footnote{A few other operators are usually considered in the effective Lagrangian associated to these LFV observables, but the corresponding effects are negligible. See Ref.~\cite{Cirigliano:2005ck} for further details.}
\be
\LL^\text{eff}_\text{LFV}=\dfrac{1}{\LLF^2}\sum_{i=1}^{5}c^{(i)}_{LL}\O_{LL}^{(i)}+\dfrac{1}{\LLF^2}\left(\sum_{j=1}^2 c_{RL}^{(j)}\O_{RL}^{(j)}+\hc\right)\,,
\label{EffLagLFV}
\ee
where the Lagrangian parameters are real coefficients\footnote{The reality of the Lagrangian parameters guarantees that no sources of CP violation are introduced beyond the SM. A justification of this approach can be found in Ref.~\cite{Paradisi:2009ey}.} of order 1 and the operators have the form\footnote{The notation chosen for the effective operators matches the one of the original MLFV paper~\cite{Cirigliano:2005ck}. It is nowadays common to adopt an other operator basis introduced in Ref.~\cite{Buchmuller:1985jz,Grzadkowski:2010es}. The link between the two bases is given by:
\be
\begin{aligned}
\O_{LL}^{(1)}&\to Q_{\varphi\ell}^{(1)}\,,\qquad\qquad
\O_{LL}^{(2)}&\to Q_{\varphi\ell}^{(3)}\,,\qquad\qquad
\O_{LL}^{(3)}&\to Q_{\ell q}^{(1)}\,,\qquad\qquad
\O_{LL}^{(4d)}&\to Q_{\ell d}\,,
\\
\O_{LL}^{(4u)}&\to Q_{\ell d}\,,\qquad\qquad
\O_{LL}^{(5)}&\to Q_{\ell q}^{(3)}\,,\qquad\qquad
\O_{RL}^{(1)}&\to Q_{eB}\,,\qquad\qquad
\O_{RL}^{(2)}&\to Q_{eW}\,.
\end{aligned}
\ee}:
\be
\begin{aligned}
\O_{LL}^{(1)}&=i\ov \ell\gamma^\mu\ell_L H^\dag D_\mu H\,,\qquad\qquad
&
\O_{LL}^{(2)}&=i\ov \ell\gamma^\mu\sigma^a\ell_L H^\dag \sigma^a D_\mu H\,,
\\
\O_{LL}^{(3)}&=\ov \ell\gamma^\mu\ell_L \ov q\gamma_\mu q_L\,,\qquad\qquad
&
\O_{LL}^{(4d)}&=\ov \ell\gamma^\mu\ell_L \ov d\gamma_\mu d_R\,,
\\
\O_{LL}^{(4u)}&=\ov \ell\gamma^\mu\ell_L \ov u\gamma_\mu u_R\,,\qquad\qquad
&
\O_{LL}^{(5)}&=\ov \ell\gamma^\mu\sigma^a\ell_L \ov q\gamma_\mu\sigma^a q_L\,,
\\
\O_{RL}^{(1)}&=g' \ov\ell H \sigma^{\mu\nu} e_R B_{\mu\nu}\,,\qquad\qquad
&
\O_{RL}^{(2)}&=g \ov\ell H \sigma^{\mu\nu}\sigma^a e_R W^a_{\mu\nu}\,.
\end{aligned}
\label{HEEffOperators}
\ee

The $\O_{LL}^{(i)}$ structures are invariant under the flavour symmetries without the necessity of introducing any spurion field, but they can only contribute to flavour conserving observables. The LFV processes aforementioned can only be described by the insertion of specific spurion combinations transforming as ${\bf 8}$ under $SU(3)_{\ell_L}$, whose flavour indices are contracted with those of the lepton bilinear $\ov{\ell}_L^i\Gamma \ell_L^j$ in $\O_{LL}^{(i)}$,
$\Gamma$ being a suitable combination of Dirac and/or Pauli matrices.
The specific spurion combinations depend on the considered model: some examples are $\g_\nu^\dag\,\g_\nu$ in MFC, $\Y_\nu \Y_\nu^\dag$ in EFCI and $\Y_\nu \Y_N^\dag \Y_N \Y_\nu^\dag$ in EFCII. Interestingly, once the spurions acquire their background values, these combinations reduce to the expressions for $\Delta$ in Eqs.~(\ref{DeltaMFC}), (\ref{DeltaEFCI}) and (\ref{DeltaEFCII}), respectively.

The $\O_{RL}^{(i)}$ operators, instead, are not invariant under the flavour symmetry $\G_L$ and require the insertion of spurion combinations transforming as $({\bf 3},\,\ov{\bf 3})$ under $SU(3)_{\ell_L}\times SU(3)_{e_R}$. The simplest combination of this kind is the charged lepton Yukawa spurion $\Y_e$, whose background value, however, is diagonal. Requiring as well that these structures describe LFV processes, it is necessary to insert more elaborated combinations: some examples are $\g_\nu^\dag\,\g_\nu\Y_e$ in MFC, $\Y_\nu \Y_\nu^\dag\Y_e$ in EFCI and $\Y_N^\dag \Y_N \Y_e$ in EFCII. Once the spurions acquire background values, these combinations reduce to $\Delta Y_e$, with the specific expression for $\Delta$ depending on the case considered.

From the previous discussion one can deduce that the relevant quantity that allows to describe LFV processes in terms of lepton masses and mixings is $\Delta$, beside the diagonal matrix $Y_e$. It is then instructive to explicitly write the expression for $\Delta$ in the three cases under consideration and distinguishing between the NO and the IO for the neutrino mass spectrum\footnote{The expression for $\Delta$ in the IO case may differ from what reported in Ref.~\cite{Cirigliano:2005ck}, due to a different definition taken for the atmospheric mass squared difference.}.

\begin{itemize}
\item[1.] Minimal Field Content $\G_L=SU(3)_{\ell_L}\times SU(3)_{e_R}$. Expliciting Eq.~(\ref{DeltaMFC}), the off-diagonal entries of $\Delta$ can be written as
\be
\begin{aligned}
\Delta_{\mu e}=&\dfrac{4\LLN^2}{v^4}\left[s_{12}c_{12}c_{23}c_{13}\left(m_{\nu_B}-m_{\nu_A}\right)+s_{23}s_{13}c_{13}e^{i\delta}\left(m_{\nu_C}-s_{12}^2m_{\nu_B}-c_{12}^2 m_{\nu_A}\right)\right]\,,\\
\Delta_{\tau e}=&\dfrac{4\LLN^2}{v^4}\left[-s_{12}c_{12}s_{23}c_{13}\left(m_{\nu_B}-m_{\nu_A}\right)+c_{23}s_{13}c_{13}e^{i\delta}\left(m_{\nu_C}-s_{12}^2m_{\nu_B}-c_{12}^2m_{\nu_A}\right)\right]\,,\\
\Delta_{\tau \mu}=&\dfrac{4\LLN^2}{v^4}\left\{s_{23}c_{23}\left[c_{13}^2m_{\nu_C}+(s_{12}^2s_{13}^2-c_{12}^2)m_{\nu_B}+(c_{12}^2s_{13}^2-s_{12}^2)m_{\nu_A}\right]+\right. \\
& \hspace{5cm}\left. +s_{12}c_{12}s_{13}\left(s_{23}^2e^{-i\delta}-c_{23}^2e^{i\delta}\right)\left(m_{\nu_B}-m_{\nu_A}\right)\right\}\,,
\end{aligned}
\label{DeltaMFCExp}
\ee
where, for brevity of notation, $s_{ij}$ and $c_{ij}$ stand for the sine and cosine of the leptonic mixing angles $\theta^\ell_{ij}$, $\delta$ stands for the Dirac CP phase $\delta^\ell_{CP}$, and a generic notation for $\hat M_\nu$ has been adopted in the definition of $\Delta$:
\be
\hat M^2_\nu\equiv\diag\left(m_{\nu_A},\,m_{\nu_B},\,m_{\nu_C}\right)\,.
\label{MnuMFC}
\ee
The three parameters $m_{\nu_{A,B,C}}$ depend on the neutrino mass ordering: for the NO case
\be
\begin{aligned}
m_{\nu_A}=0\,,\quad
m_{\nu_B}=\Dmsol\,,\quad
m_{\nu_C}=\Dmatm\,,
\end{aligned}
\ee
and for the IO case
\be
\begin{aligned}
m_{\nu_A}=\Dmatm-\Dmsol\,,\quad
m_{\nu_B}=\Dmatm\,,\quad
m_{\nu_C}=0\,.
\end{aligned}
\ee
Notice that there is no dependence on the lightest neutrino mass in these expressions. This has an interesting consequence because $\Delta_{i\neq j}$ are completely fixed, apart for the common scale $\LLN$.

\item[2.] Extended Field Content I) $\G_L=SU(3)_{\ell_L}\times SU(3)_{e_R}\times SO(3)_{N_R}\times CP$. From Eqs.~(\ref{DeltaEFCI}), one gets the following explicit expressions for the off-diagonal entries of $\Delta$:
\be
\begin{aligned}
\Delta_{\mu e}=&\frac{2\muLN}{v^2}\left[s_{12}c_{12}c_{23}c_{13}\left(m_{\nu_B}-m_{\nu_A}\right)+s_{23}s_{13}c_{13}e^{i\delta}\left(e^{-2i\delta}m_{\nu_C}-s_{12}^2m_{\nu_B}-c_{12}^2 m_{\nu_A}\right)\right]\,,\\
\Delta_{\tau e}=&\frac{2\muLN}{v^2}\left[-s_{12}c_{12}s_{23}c_{13}\left(m_{\nu_B}-m_{\nu_A}\right)+c_{23}s_{13}c_{13}e^{i\delta}\left(e^{-2i\delta}m_{\nu_C}-s_{12}^2m_{\nu_B}-c_{12}^2m_{\nu_A}\right)\right]\,,\\
\Delta_{\tau \mu}=&\frac{2\muLN}{v^2}\left\{s_{23}c_{23}\left(c_{13}^2m_{\nu_C}-c_{12}^2m_{\nu_B}-s_{12}^2m_{\nu_A}\right)+\right. \\
& \left. +s_{12}c_{12}s_{13}e^{i\delta}(s_{23}^2-c_{23}^2)\left(m_{\nu_B}-m_{\nu_A}\right)+s_{23}c_{23}s^2_{13}e^{2i\delta}\left(s_{12}^2m_{\nu_B}+c_{12}^2 m_{\nu_A}\right)\right\}\,,
\end{aligned}
\label{DeltaEFCIExp}
\ee
where a generic notation -different from the one in the MFC case- for $\hat M_\nu$ has been adopted:
\be
\hat M_\nu\equiv\diag\left(m_{\nu_A},\,m_{\nu_B},\,m_{\nu_C}\right)\,.
\label{MnuEFCI}
\ee
The three parameters $m_{\nu_{A,B,C}}$ are now defined by
\be
\begin{aligned}
m_{\nu_A}=m_{\nu_1}\,,\quad
m_{\nu_B}=e^{i\alpha_{21}}\sqrt{\Dmsol+m_{\nu_1}^2}\,,\quad
m_{\nu_C}=e^{i\alpha_{31}}\sqrt{\Dmatm+m_{\nu_1}^2}\,,
\end{aligned}
\ee
for the NO case, $m_{\nu_1}<m_{\nu_2}<m_{\nu_3}$,
and by
\be
\begin{aligned}
m_{\nu_A}=\sqrt{\Dmatm-\Dmsol+m_{\nu_3}^2}\,,\quad
m_{\nu_B}=e^{i\alpha_{21}}\sqrt{\Dmatm+m_{\nu_3}^2}\,,\quad
m_{\nu_C}=e^{i\alpha_{31}}m_{\nu_3}\,,
\end{aligned}
\ee
for the IO case, $m_{\nu_3}<m_{\nu_1}<m_{\nu_2}$.

The hypothesis of CP conservations fixes the Dirac and Majorana CP phases to be $\delta=\{0,\pi\}$ and $\alpha_{21,31}=0$ in these expressions. Indeed, while $\Delta_{ij}$ would be real even for $\alpha_{21,31}=\pi$ and therefore no CPV process would be described with $\Delta$ insertions, $Y_\nu$ would be complex and then it would not be possible to express the spurions insertions in Eq.~(\ref{FCCombinations}) in terms of low-energy parameters, losing the predictivity power of MLFV.

In the strong hierarchical limit, $m_{\nu_1}\ll m_{\nu_2}<m_{\nu_3}$ in the NO case and
$m_{\nu_3}\ll m_{\nu_1}<m_{\nu_2}$ in the IO one,
and setting the lightest neutrino mass to zero,
the expressions for $m_{\nu_{A,B,C}}$ reduce to the square root
of those for the MFC case, as can be deduced comparing
Eqs.~(\ref{MnuMFC}) and (\ref{MnuEFCI}), and
the results for $\Delta_{i\neq j}$ get simplified.
Also in this case, only one parameter remains free,
that is the LNV scale $\muLN$.

When the neutrino mass hierarchy is milder or the eigenvalues are almost degenerate, the lightest neutrino mass cannot be neglected and represents a second free parameters of $\Delta_{i\neq j}$, besides $\muLN$.

\item[3.] Extended Field Content II) $\G_L=SU(3)_{\ell_L+N_R}\times SU(3)_{e_R}$. The expressions for the off-diagonal entries of $\Delta$ that follow from Eqs.~(\ref{DeltaEFCII}) can be obtained from the expressions in Eq.~(\ref{DeltaMFCExp}), by substituting
\be
\dfrac{4\LLN^2}{v^4}\to\dfrac{v^4}{4\muLN^2}
\ee
and taking the following notation for $\hat M_\nu$:
\be
\hat M^{-2}_\nu\equiv\diag\left(m_{\nu_A},\,m_{\nu_B},\,m_{\nu_C}\right)\,,
\ee
with $m_{\nu_{A,B,C}}$ given by
\be
\begin{aligned}
m_{\nu_A}=\frac{1}{m^2_{\nu_1}}\,,\quad
m_{\nu_B}=\frac{1}{\Dmsol+m_{\nu_1}^2}\,,\quad
m_{\nu_C}=\frac{1}{\Dmatm+m_{\nu_1}^2}\,,
\end{aligned}
\label{MassesEFCII}
\ee
for the NO case, and
\be
\begin{aligned}
m_{\nu_A}=\frac{1}{\Dmatm-\Dmsol+m_{\nu_3}^2}\,,\quad
m_{\nu_B}=\frac{1}{\Dmatm+m_{\nu_3}^2}\,,\quad
m_{\nu_C}=\frac{1}{m_{\nu_3}}\,,
\end{aligned}
\ee
for the IO case.

The limits for the lightest neutrino mass being zero are not well defined
for this case, as it would lead to an infinity in the expressions for
$\Delta_{i\neq j}$. Differently from the other two cases, only a moderate
neutrino mass hierarchy is then allowed. Finally, these expressions depend
on two free parameters, the lightest neutrino mass and the LNV scale $\muLN$.

\end{itemize}

\subsection{Rare Radiative Leptonic Decays and Conversion in Nuclei}

In the formalism of the effective Lagrangian reported in the Eq.~(\ref{EffLagLFV}), the Beyond SM (BSM) contributions to the branching ratio of leptonic  radiative rare decays are given by
\be
B_{\ell_i\to\ell_j\gamma}\equiv\dfrac{\Gamma(\ell_i\to\ell_j\gamma)}{\Gamma(\ell_i\to\ell_j\nu_i\ov\nu_j)}=
384\pi^2e^2\dfrac{v^4}{4\LLF^4}\left|\Delta_{ij}\right|^2\left|c_{RL}^{(2)}-c_{RL}^{(1)}\right|^2\,,
\label{BRadiativeGen}
\ee
being $e$ the electric charge, and where the corrections of the Wilson coefficient due to the electroweak renormalisation from the scale of NP down to the mass scale of the interested lepton\cite{Pruna:2014asa,Crivellin:2017rmk} have been neglected, and the limit $m_{\ell_j}\ll m_{\ell_i}$ has been taken.

The same contributions to the branching ratio for $\mu\to e$ conversion in a generic nucleus of mass number $A$ read
\be
\begin{split}
B_{\mu\to e}^A=&\dfrac{32 G_F^2m_\mu^5}{\Gamma_\text{capt}^A}\dfrac{v^4}{4\LLF^4}\left|\Delta_{\mu e}\right|^2\left|\left(\left(\dfrac{1}{4}-s_w^2\right)V^{(p)}-\dfrac{1}{4}V^{(n)}\right)\left(c_{LL}^{(1)}+c_{LL}^{(2)}\right)+\right.\\
&+\dfrac{3}{2}\left(V^{(p)}+V^{(n)}\right)c_{LL}^{(3)}+\left(V^{(p)}+\dfrac{1}{2}V^{(n)}\right)c_{LL}^{(4u)}+\left(\dfrac{1}{2}V^{(p)}+V^{(n)}\right)c_{LL}^{(4d)}+\\
&\left.+\dfrac{1}{2}\left(-V^{(p)}+V^{(n)}\right)c_{LL}^{(5)}-\dfrac{eD_A}{4}\left(c_{RL}^{(2)}-c_{RL}^{(1)}\right)^*\right|^2\,,
\end{split}
\label{BConvGen}
\ee
where $s_W\equiv \sin\theta_W=0.23$, $V^{(p)}$, $V^{(n)}$ and $D$ are dimensionless nucleus-dependent overlap integrals that can be found in Tab.~\ref{NuclearInts} for Aluminium and Gold, that also contains the numerical values for decay rate of the muon capture, which has been used to normalise the decay rate for the $\mu\to e$ conversion.

\begin{table}[htp]
\begin{center}
\begin{tabular}{c|c|c|c|c|}
	& $V^{(p)}$	& $V^{(p)}$	& $D$		& $\Gamma_\text{capt}\,(10^{6}\,\text{s}^{-1})$\\[2mm]
\hline
&&&&\\
Au	& $0.0974$	& $0.146$		& $0.189$		& $13.07$\\[2mm]
Al	& $0.0161$	& $0.0173$	& $0.0362$	& $0.7054$ \\[2mm]
\hline
\end{tabular}
\end{center}
\caption{\footnotesize\it Reference values for nuclear overlap integrals and capture rates from Ref.~\cite{Kitano:2002mt}.}
\label{NuclearInts}
\end{table}

The experimental bounds on these processes that will be considered in the numerical analysis are the following:
\be
\begin{gathered}
B_{\mu\to e\gamma}<5.7\times 10^{-13}\text{~\cite{Adam:2013mnn}}\,\, (6\times 10^{-14}\text{~\cite{Baldini:2013ke}})\,,\\
B_{\tau\to e\gamma}<5.2\times 10^{-8}\text{~\cite{Aubert:2009ag}}\,\,(10^{-9}\div 10^{-10}\text{~\cite{Hayasaka:2013dsa}})\,,\\
B_{\tau\to \mu\gamma}<2.5\times 10^{-7}\text{~\cite{Aubert:2009ag}}\,\,(10^{-8}\div 10^{-9}\text{~\cite{Hayasaka:2013dsa}})\,,\\
B^\text{Au}_{\mu\to e}<7\times 10^{-13}\text{~\cite{Bertl:2006up}}\,,\\
B^\text{Al}_{\mu\to e}<6\times 10^{-17}\text{~\cite{Kuno:2013mha,Abrams:2012er}}\,,
\end{gathered}
\label{ExperimentalValuesBRs}
\ee
where the values in the brackets and the bound on $B^\text{Al}_{\mu\to e}$ refer to future expected sensitivities.

\subsubsection{Bounds on the LFV Scale}

The bounds on the LNV scales, determined in Eqs.~(\ref{LNVScaleBoundMFC}), (\ref{LNVScaleBoundEFCI}) and (\ref{LNVScaleBoundEFCII}), can be translated into bounds on the LFV scale when considering the experimental limits in the rare processes introduced above. Indeed, after substituting the expressions for $\Delta$, defined in Eqs.~(\ref{DeltaMFC}), (\ref{DeltaEFCI}) and (\ref{DeltaEFCII}), into the Eqs.~(\ref{BRadiativeGen}) and (\ref{BConvGen}), one can rewrite these expressions extracting the dependence on the NP scales:
\be
\begin{cases}
B_{\ell_i\to\ell_j(\gamma)}\equiv\left(\dfrac{\LLN}{\LLF}\right)^4 \widetilde B_{\ell_i\to\ell_j(\gamma)}\big[c_i\big]\,,\qquad\qquad&\text{for the MFC case}\\[3mm]
B_{\ell_i\to\ell_j(\gamma)}\equiv\left(\dfrac{v\muLN}{\LLF^2}\right)^2 \widetilde B_{\ell_i\to\ell_j(\gamma)}\left[m_\nu^\text{lightest},\, c_i\right]\,,\qquad\qquad&\text{for the EFCI case}\\[3mm]
B_{\ell_i\to\ell_j(\gamma)}\equiv\left(\dfrac{v^2}{\muLN\LLF}\right)^4 \widetilde B_{\ell_i\to\ell_j(\gamma)}\left[m_\nu^\text{lightest},\, c_i\right]\,,\qquad\qquad&\text{for the EFCII case}
\end{cases}
\label{BrRedefinition}
\ee
where the square brackets list the free parameters, that is the lightest neutrino mass (only for the EFCI and EFCII cases) and the effective Lagrangian parameters $c_i$.

\begin{figure}[htbp]
\begin{center}
\includegraphics[width=0.49\textwidth]{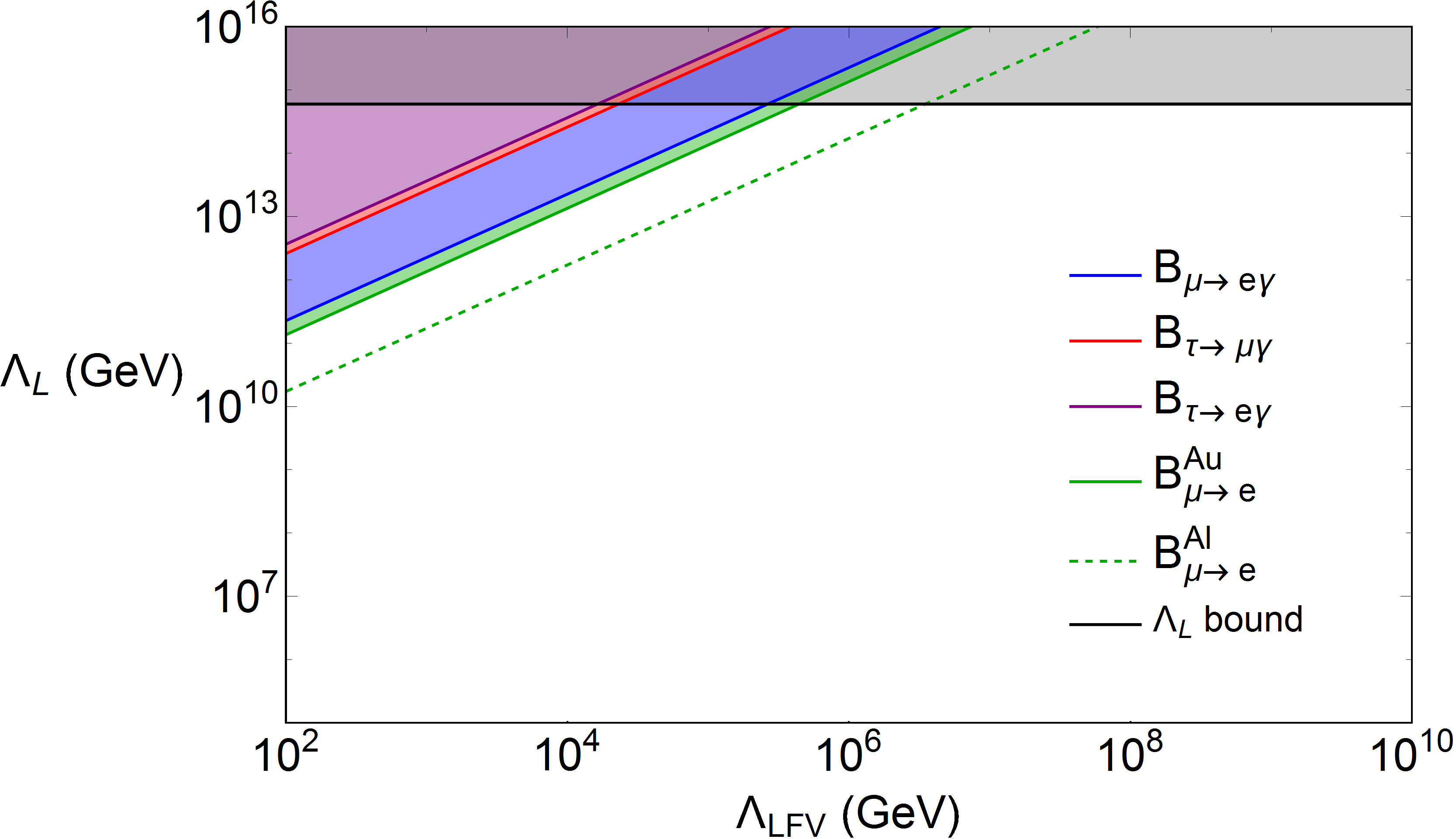}
\includegraphics[width=0.49\textwidth]{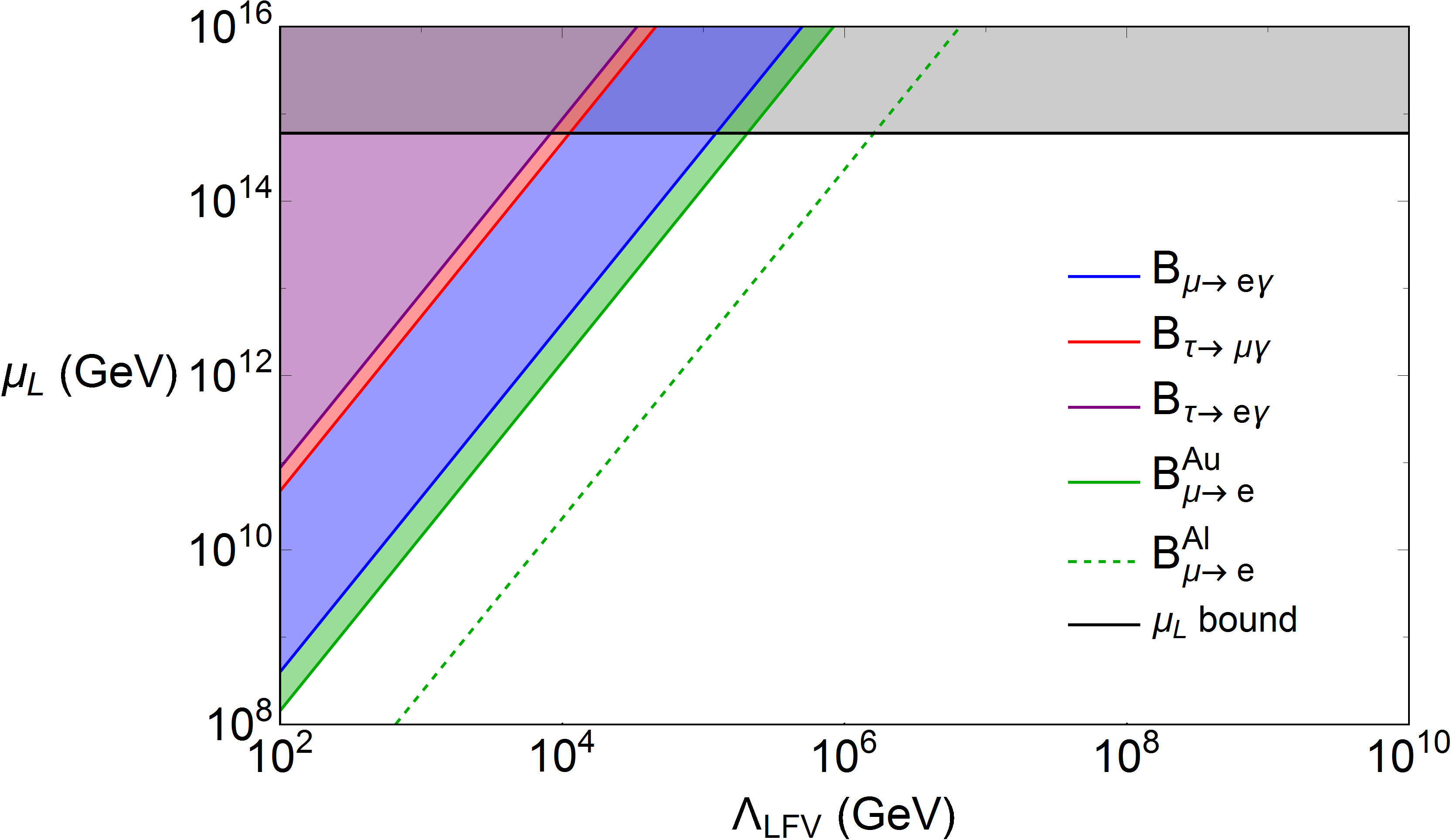}
\includegraphics[width=0.49\textwidth]{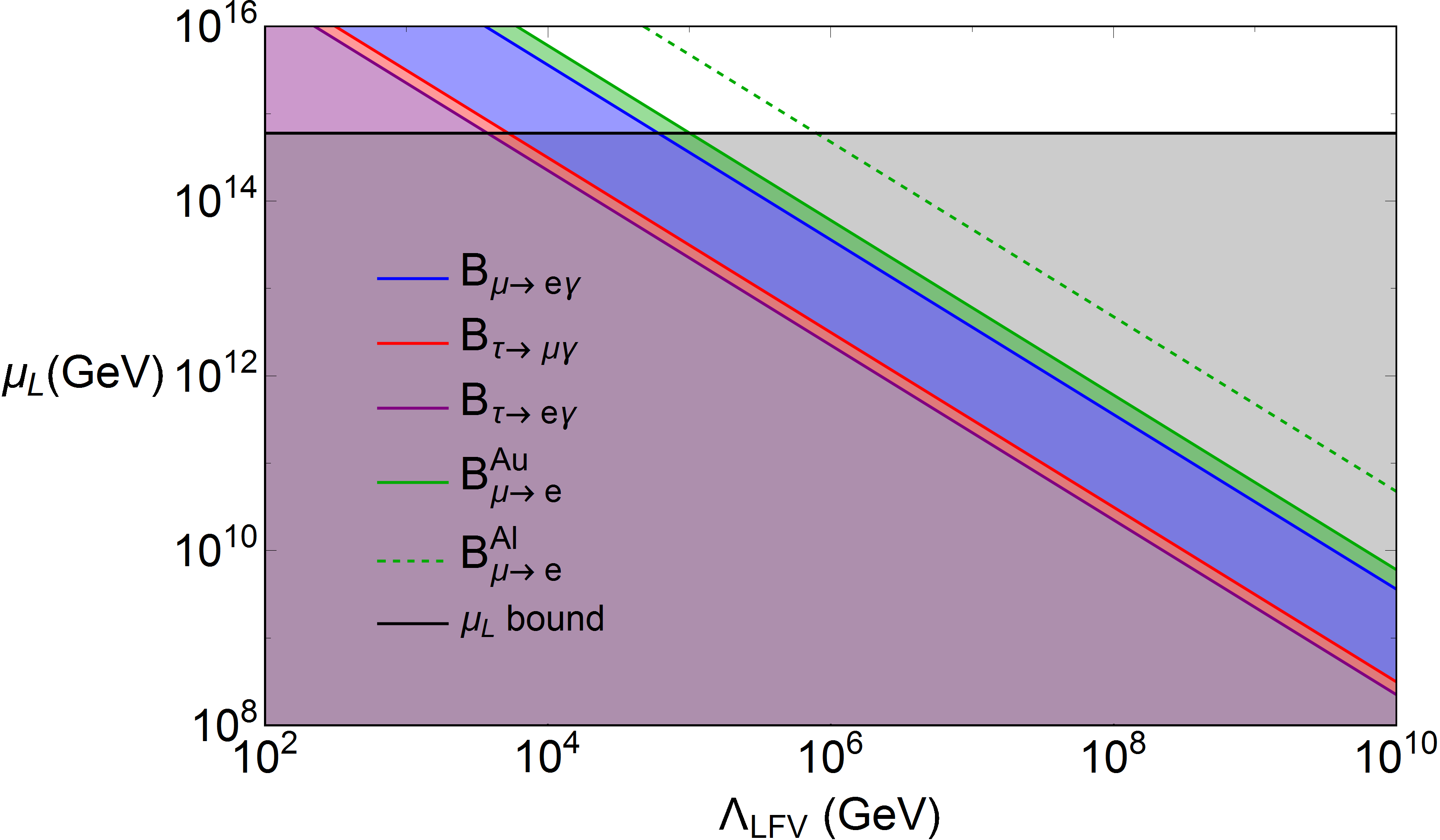}
\caption{\footnotesize\it
  Parameter space for the LFV and LNV scales constrained by requiring perturbativity
    of the spurion backgrounds and by the present experimental bounds on
    $\mu\to e $ conversion in gold (in green),
    $BR(\mu\to e \gamma)$ (in blue), $BR(\tau\to \mu \gamma)$ (in red), and
    $BR(\tau\to e \gamma)$ (in purple).
    Taking into account the expected future sensitivity on $BR(\mu\to e \gamma)$
    would not restrict further the parameter space in the case of a negative result:
    the prospective bound would almost coincide with the bound
    from the negative search for $\mu\to e$ conversion in gold, $BR(\mu\to e)$.
    However, with the planned significant
    increase (by more than 4 orders of magnitude) of the sensitivity
    to the relative rate of $\mu\to e$ conversion in aluminium
    it would be possible to probe considerably larger fraction of the
    parameter space of interest: the corresponding bound is drown as the green dashed line.
    The grey region are excluded areas from the constraints on the LNV scale,
    Eqs.~(\ref{LNVScaleBoundMFC}), (\ref{LNVScaleBoundEFCI}), and (\ref{LNVScaleBoundEFCII}).
    The left, middle and right panels correspond to the MFC, EFCI and EFCII cases, respectively.
    The border lines are obtained taking as input data the best fit values for the oscillation parameters
    listed in Tab.~\ref{TableOscFit} and the nuclear quantities in Tab.~\ref{NuclearInts}.
    The Dirac CP phase for the EFCI plot is set equal to $\pi$, while the Majorana are set to $0$,
    in order to minimise the excluded region of the parameter space.
    For the EFCI and EFCII cases, a quasi-degenerate
    neutrino mass spectrum with $m_\nu^\text{lightest}=0.1\eV$ has been assumed, which also
    minimised the excluded areas. In all the cases, the Lagrangian coefficients
    have been fixed in a democratic way not to favour
  any specific operator contribution: $c_{LL}^{(1)}+c_{LL}^{(2)}=1=c_{LL}^{(3)}=c_{LL}^{(4u)}=c_{LL}^{(4d)}=
  c_{LL}^{(5)}=c_{RL}^{(2)}-c_{RL}^{(1)}$.}
\label{ScalesPlots}
\end{center}
\end{figure}

The numerical analysis reveals that the strongest bounds on the $\LLF$ comes from the data on $\mu\to e$ conversion in gold, although similar results are provided by the data on leptonic radiative rare decays. The corresponding parameter space is shown in Fig.~\ref{ScalesPlots}, obtained taking the best fit values for the quantities in Tab.~\ref{TableOscFit} (for the EFCI case, the Dirac CP phase can only acquire two values, $0$ and $\pi$) and the data from Tab.~\ref{NuclearInts}. Although these plots have been generated for the NO neutrino spectrum, they hold for the IO case as well, as no difference is appreciable. On the other hand, a dependence on the strength of the splitting between neutrino masses can be found for the EFC scenarios: the plots reported here illustrate the almost degenerate case, where the lightest neutrino mass is taken to be $\O(0.1\eV)$; stronger hierarchies result in a more constrained parameter space. Finally, the plot for EFCI refers to $\delta^\ell_{CP}=\pi$, but the other case with
$\delta^\ell_{CP}=0$ is almost indistinguishable.

The upper bound on $\LLN$ for the MFC case reduce the parameter space, although it cannot be translated into upper
bounds on $\LLF$: larger $\LLF$ simply further suppresses the expected values for the branching ratios of the observables considered. 
Moreover, no lower bound can be drown: requiring to close the experimental bound for the $\mu\to e$ conversion,
small $\LLF$ requires small $\LLN$, leading at the same time to tune $g_\nu$ to small values, in order to reproduce the correct masses for the light active neutrinos, see Eq.~(\ref{BackgroudSpurionsMFC}). The same occurs for EFCI, for $\muLN$ and $Y_\nu$, although, in this case, this can be well justified considering the additional Abelian symmetries appearing in Eq.~(\ref{FullGLSS}), as discussed in Ref.~\cite{Alonso:2011jd}.
When considering the EFCII case, the lower bound on $\LLN$ removes a large part of the parameter space, but does not translate into a lower bound on $\LLF$: for example, for $\LLN$ at its lower bound in Eq.~(\ref{LNVScaleBoundEFCII}), $\LLF$ must be larger than $10^5\GeV$ in order to satisfy to the present bounds on $B_{\mu\to e}^\text{Au}$; however, for larger values of $\LLN$, $\LLF$ can be smaller, down to the TeV scale for $\LLN\sim 10^{17}\GeV$, although in this case a tuning on $|Y_N|$ is necessary in order to reproduce correctly the lightness of the active neutrino masses.

The absence of evidence of NP in direct and indirect searches at colliders and low-energy experiments suggests that NP leading to LFV should be heavier than a few TeV. In the optimistic scenario that NP is just behind the corner and waiting to be discovered in the near future, an indication of the LNV scale
could be extracted from the plots in Fig.~\ref{ScalesPlots}. Indeed, if $\mu\to e$ conversion in nuclei is observed, $\LLF\sim 10^3\div10^4\GeV$ will lead to $\LLN\sim10^{12}\div 10^{13}\GeV$ for MFC, $\muLN\sim10^{9}\div 10^{10}\GeV$ for EFCI, and $\muLN\sim10^{16}\div10^{17}\GeV$ for EFCII. In the EFC scenarios, the LNV scale is associated to the masses of the RH neutrinos, that therefore turn out to be much heavier than the energies reachable at present and future colliders. An exception is the case where additional Abelian factors are considered in the flavour symmetry that allows to separate the LNV scale and the RH neutrino masses~\cite{Alonso:2011jd}: this opens the possibility of producing sterile neutrinos at colliders and then of studying their interactions in direct searches.

\subsubsection{Ratios of Branching Ratios}

The information encoded in Eq.~(\ref{BrRedefinition}) are not limited to the scales of LFV and LNV.
Studying the ratios of branching ratios between the different processes reveals
characteristic features that may help to disentangle
the different versions of MLFV.
To shorten the notation,
\be
R^{t\to s\gamma}_{i\to j\gamma}\equiv \dfrac{\widetilde B_{\ell_t\to\ell_s\gamma}}{\widetilde B_{\ell_i\to\ell_j\gamma}}\,,
\ee
will be adopted in the analysis that follows. These observables do not depend on the LFV and LNV scales, nor on the Lagrangian coefficients. They are sensible to the neutrino oscillation parameters and, for the EFC cases, to the mass of the lightest active neutrino. For MFC, they do not even depend on $m_\nu^\text{lightest}$: although the corresponding plots only contain points along an horizontal line, they will be reported in the next subsections in order to facilitate the comparison with the other cases.

The two branching ratios with the best present sensitivities, the one for $\mu\to e$ conversion in nuclei and the one for $\mu\to e\gamma$, have the same dependence on
$\Delta_{\mu e}$ and therefore their ratio is not sensitive to the charged lepton and neutrino masses and to the neutrino mixing.
Instead, as pointed out in Ref.~\cite{Cirigliano:2006su}, this ratio may be sensitive to
the chirality of the effective operators contributing to these observables.
The comparison between Eqs.~(\ref{BRadiativeGen}) and (\ref{BConvGen})
shows that only $B^A_{\mu\to e}$ is sensitive to $\O^{(i)}_{LL}$, and thus any deviation from
\be
\dfrac{B_{\mu\to e}^A}{B_{\mu\to e\gamma}}=\pi D_A^2
\label{BRatioConv}
\ee
would be a signal of this set of operators.

In the scatter plots that follow, neutrino oscillation parameters are
taken from Tab.~\ref{TableOscFit} as random values inside their
$2\sigma$ error bands. The lightest neutrino mass is taken in the
range $m_\nu^\text{lightest}\subset[0.001,\,0.1]\eV$ and the results for the NO and IO spectra
are shown with different colours. In these figures, the density of the points
should not be interpreted as related to the likelihood of differently populated regions
of the parameter space.

\subsubsection*{\boldmath $R^{\mu\to e\gamma}_{\tau\to \mu\gamma}$}

\begin{figure}[h!]
    \centering
    \subfigure[MFC]{
    \includegraphics[width=0.47\textwidth]{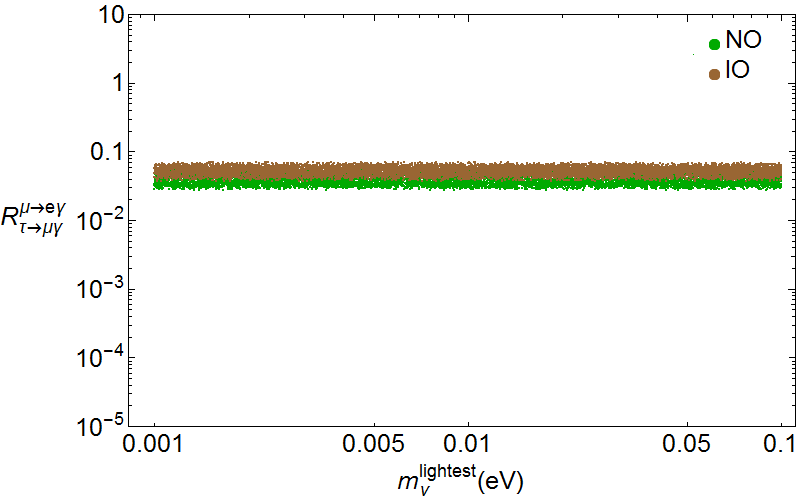}
    \label{fig:BmuetaumumlMin}}
    \subfigure[EFCI]{
    \includegraphics[width=0.47\textwidth]{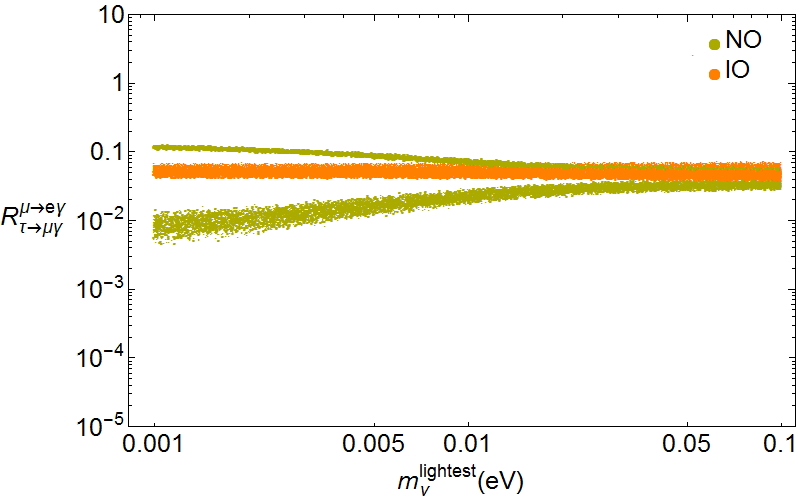}
    \label{fig:BmuetaumumlO3}}
    \subfigure[EFCII]{
    \includegraphics[width=0.47\textwidth]{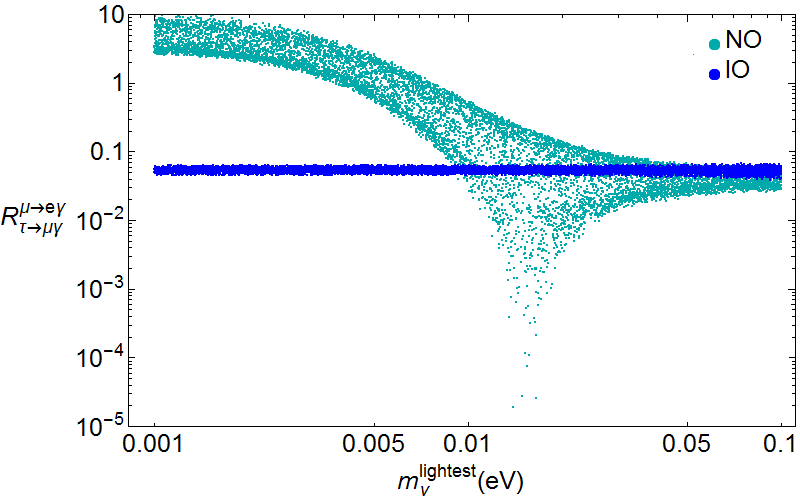}
    \label{fig:BmuetaumumlSUD}}
    \subfigure[All Cases]{
    \includegraphics[width=0.47\textwidth]{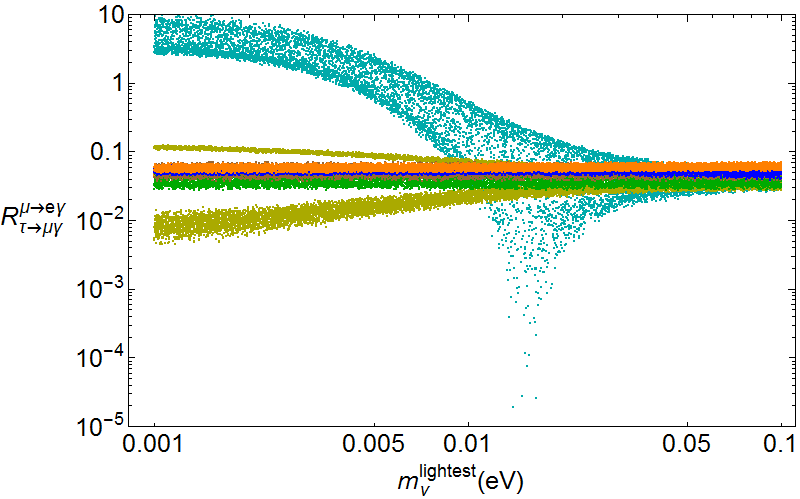}
    \label{fig:BmuetaumumlAll}}
\caption{\footnotesize\it $R^{\mu\to e\gamma}_{\tau\to \mu\gamma}$ for the MFC, EFCI and EFCII from upper left to lower left. Lower right reports the previous plots altogether. Colour codes can be read directly on each plot.}
\label{fig:Bmuetaumumlact}
\end{figure}

In the upper left, upper right and lower left panes in Fig.~\ref{fig:Bmuetaumumlact}, the results are reported for the ratio of the branching ratios of the $\mu\to e\gamma$ and $\tau\to \mu\gamma$ decays for the MFC, EFCI and EFCII cases,
respectively. Figure ~\ref{fig:BmuetaumumlAll} is a summarising figure where all the three plots are shown together to facilitate the comparison and to make clearer the non-overlapping areas.

As Fig.~\ref{fig:BmuetaumumlMin} shows, $R^{\mu\to e\gamma}_{\tau\to \mu\gamma}$ is independent of the lightest neutrino mass. The two sets of points corresponding to NO and IO spectra almost overlap, making it very hard
to distinguish between the two neutrino mass orderings.

In Fig.~\ref{fig:BmuetaumumlO3},
the dependence on $m_\nu^\text{lightest}$ can be slightly
appreciated and the predictions for two mass orderings
do not overlap when the spectrum is hierarchical.
In the NO case there are two branches associated
with the two values of $\delta^\ell_{CP}$:
the values associated with the $\delta^\ell_{CP}=0$-branch are very
close to those for the IO spectrum and correspond to
the positive sum of the two terms on the right-hand side of
Eq.~(\ref{DeltaEFCIExp});  the values associated with the
$\delta^\ell_{CP}=\pi$-branch are smaller by
about one order of magnitude, which reflects a partial
cancellation between the two terms in the right-hand side of
Eq.~(\ref{DeltaEFCIExp}). In the IO case there is only one branch
 because the first term on the right-hand side of Eq.~(\ref{DeltaEFCIExp})
is dominant.

As  Fig.~\ref{fig:BmuetaumumlSUD} shows, the points for the two mass orderings
overlap in the quasi-degenerate limit down to masses of about $0.05\eV$.
However,  they show different profiles in the hierarchical limit.
In the IO case the ratio of branching ratios under discussion
is almost constant with $m_\nu^\text{lightest}$.
In the NO case the ratio $R^{\mu\to e\gamma}_{\tau\to \mu\gamma}$
can be as small as few $\times 10^{-4}$ at $\sim 0.012\eV$, while for $m_{\nu 1} < 0.01\eV$
the ratio is $R^{\mu\to e\gamma}_{\tau\to \mu\gamma} > 1$.
As discussed in Ref.~\cite{Alonso:2011jd}, this can be understood
from Eqs.~(\ref{DeltaMFCExp})
and (\ref{MassesEFCII}): in the NO case and strong mass hierarchy,
the dominant contribution
is proportional to $1/m_{\nu_1}$ and therefore
$R^{\mu\to e\gamma}_{\tau\to \mu\gamma}$ gets enhanced;
while when the spectrum is almost degenerate and in the IO case,
the dominant contribution
is suppressed by the sine of the reactor angle and the dependence on
the lightest neutrino mass is negligible. 

In Fig.~\ref{fig:BmuetaumumlAll},
where the three cases are shown altogether,
it can be seen that all the cases overlap for the IO spectrum
and in the quasi-degenerate limit for  the NO spectrum,
predicting  $R^{\mu\to e\gamma}_{\tau\to \mu\gamma} \cong 0.02\div0.07$.
When the mass spectrum is of NO type and hierarchical,
the ratio spans values from $0.004$ to $10$. Interestingly, if this ratio is observed to be larger than $0.1$,
or smaller than $0.004$, then only the EFCII with NO spectrum
can explain it. Notice that, given the current limits on $B_{\mu\to e \gamma}$, values smaller than $\sim 6\times 10^{-4}$ would be testable in the future planned experiments searching for $\tau\to \mu\gamma$. 

\subsubsection*{\boldmath $R^{\mu\to e\gamma}_{\tau\to e\gamma}$}

\begin{figure}[h!]
    \centering
    \subfigure[MFC]{
    \includegraphics[width=0.47\textwidth]{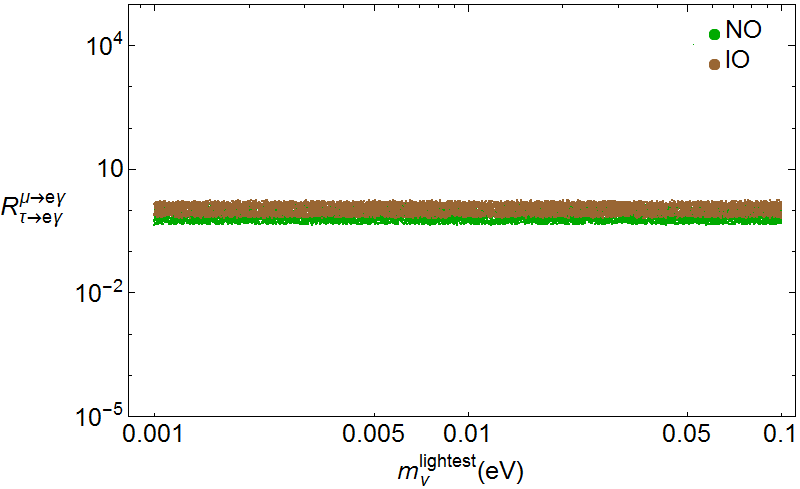}
    \label{fig:BmuetauemlMin}}
    \subfigure[EFCI]{
    \includegraphics[width=0.47\textwidth]{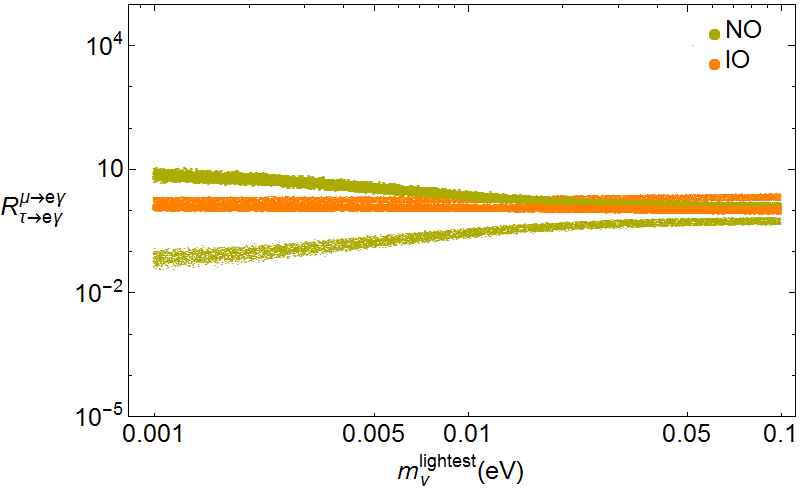}
    \label{fig:BmuetauemlO3}}
    \subfigure[EFCII]{
    \includegraphics[width=0.47\textwidth]{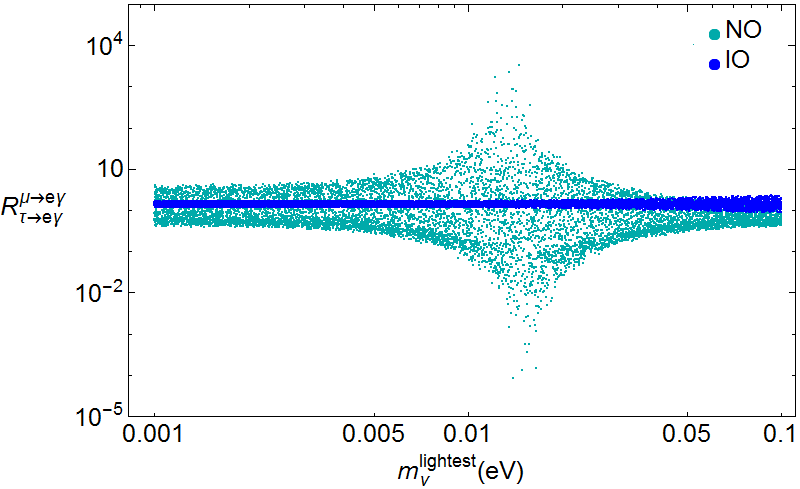}
    \label{fig:BmuetauemlSUD}}
    \subfigure[All Cases]{
    \includegraphics[width=0.47\textwidth]{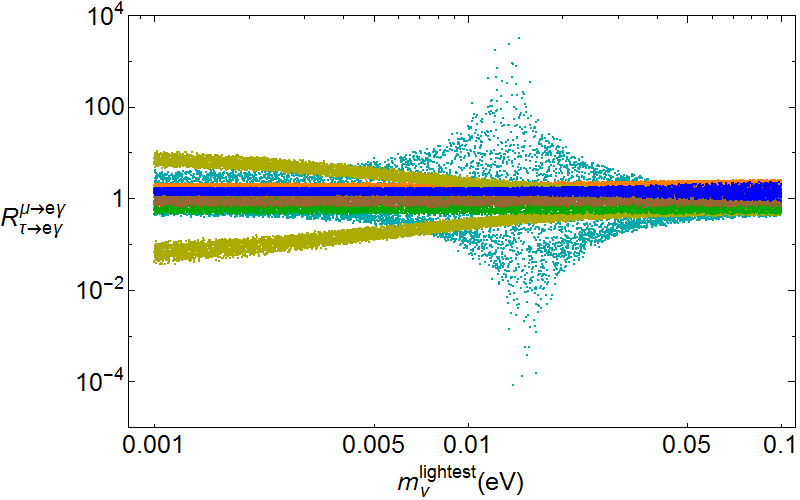}
    \label{fig:BmuetauemlSUD&O3}}
\caption{\footnotesize\it $R^{\mu\to e\gamma}_{\tau\to e\gamma}$ for the MFC, EFCI and EFCII from upper left to lower left. Lower right reports the previous plots altogether. Colour codes can be read directly on each plot.}
\label{fig:Bmuetauemlall}
\end{figure}

The ratio $R^{\mu\to e\gamma}_{\tau\to e\gamma}$
exhibits features which are very similar to those
of the ratio $R^{\mu\to e\gamma}_{\tau\to \mu\gamma}$.
Figs.~\ref{fig:BmuetauemlMin} and \ref{fig:BmuetauemlO3} are very similar
to Figs.~\ref{fig:BmuetaumumlMin} and \ref{fig:BmuetaumumlO3}: the profiles
of the points are the same, only the area spanned is different,
as indeed $R^{\mu\to e\gamma}_{\tau\to e\gamma}$ is predicted to be by almost
one order of magnitude larger than $R^{\mu\to e\gamma}_{\tau\to \mu\gamma}$.
Similar conclusions, however, apply. Fig.~\ref{fig:BmuetauemlSUD}, instead,
shows an interesting difference with respect to its
sibling Fig.~\ref{fig:BmuetaumumlSUD}:
the IO and the NO points cover almost the same nearly
horizontal area both for quasi-degenerate masses and for a
hierarchical mass spectrum, the NO region being slightly wider.
Only for values of the lightest neutrino mass between $0.01\eV$ and $0.02\eV$,
there could be an enhancement or a suppression of
$R^{\mu\to e\gamma}_{\tau\to e\gamma}$ in the EFCII case. This is
a distinctive feature
that could allow to disentangle EFCII from the other cases:
values of $R^{\mu\to e\gamma}_{\tau\to e\gamma}$ larger than 10 or smaller
than $0.04$ can only be explained by a NO neutrino spectrum in the case of
EFCII. Notice that, given the current limits on $B_{\mu\to e \gamma}$, values smaller than $0.006$ would be testable in the future planned experiments searching for $\tau\to e\gamma$. 

\subsubsection*{\boldmath $R^{\tau\to e\gamma}_{\tau\to \mu\gamma}$}

\begin{figure}[h!]
    \centering
    \subfigure[MFC]{
    \includegraphics[width=0.47\textwidth]{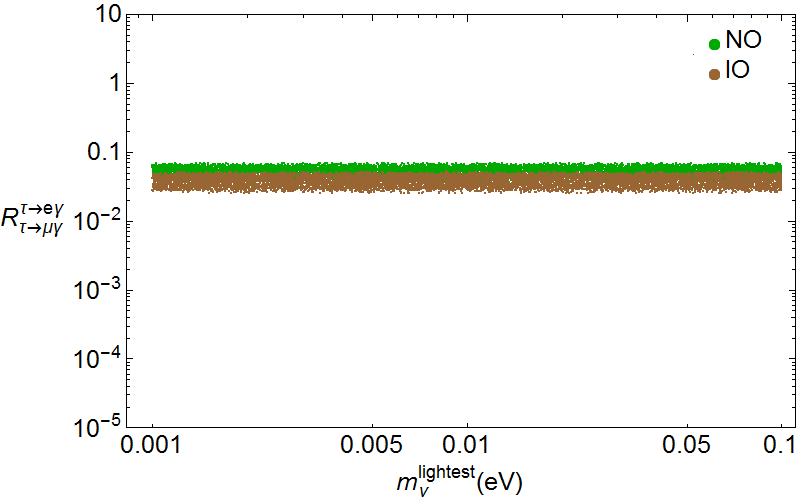}
    \label{fig:BtaumutauemlMin}}
    \subfigure[EFCI]{
    \includegraphics[width=0.47\textwidth]{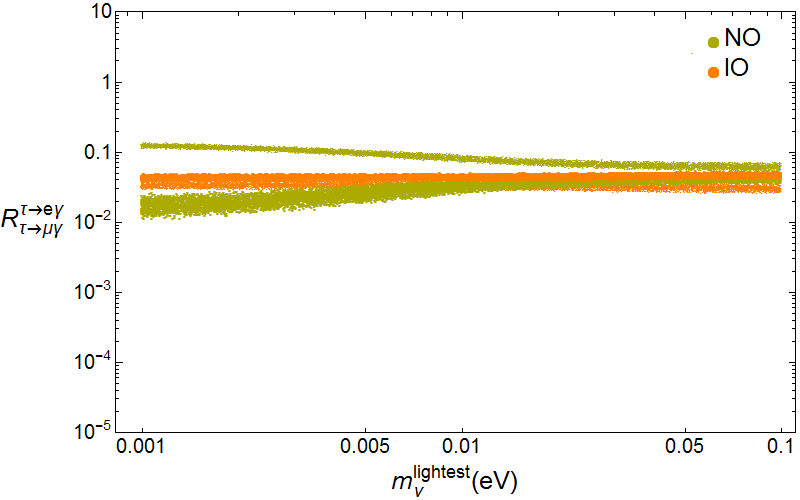}
    \label{fig:BtaumutauemlO3}}
    \subfigure[EFCII]{
    \includegraphics[width=0.47\textwidth]{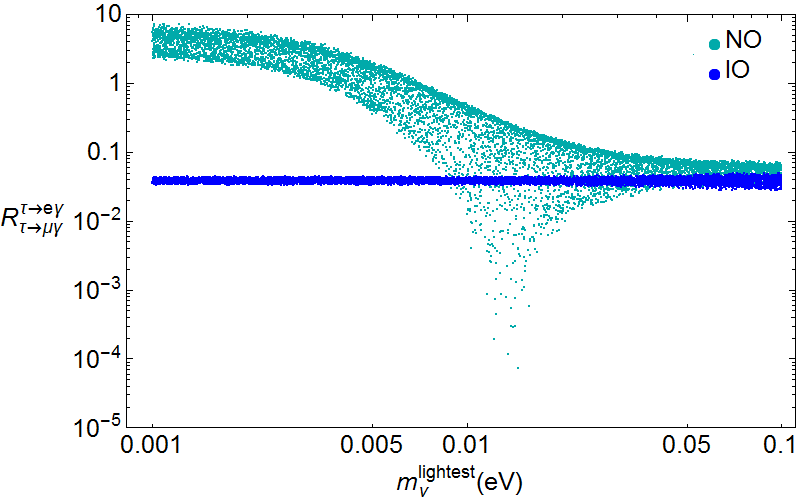}
    \label{fig:BtaumutauemlSUD}}
    \subfigure[All Cases]{
    \includegraphics[width=0.47\textwidth]{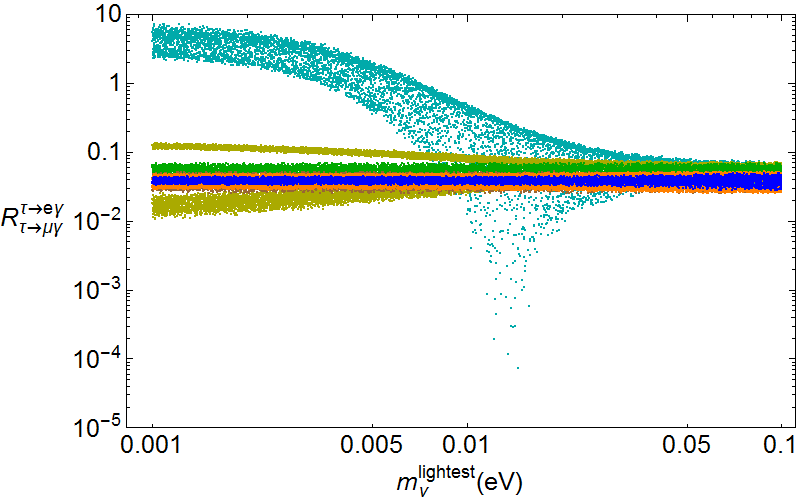}
    \label{fig:BtaumutauemlSUD&O3}}
\caption{\footnotesize\it $R^{\tau\to e\gamma}_{\tau\to \mu\gamma}$ for the MFC, EFCI and EFCII from upper left to lower left. Lower right reports the previous plots altogether. Colour codes can be read directly on each plot.}
\label{fig:Btaumutaueml}
\end{figure}

The ratio $R^{\tau\to e\gamma}_{\tau\to \mu\gamma}$ is almost indistinguishable
form the ratio $R^{\mu\to e\gamma}_{\tau\to \mu\gamma}$
except for the EFCII case with NO neutrino mass spectrum.
For the other cases the conclusions  for  $R^{\tau\to e\gamma}_{\tau\to \mu\gamma}$
are almost the same as the conclusions reached for
 $R^{\mu\to e\gamma}_{\tau\to \mu\gamma}$.
 One can see that values for
$R^{\tau\to e\gamma}_{\tau\to \mu\gamma}$ smaller than $0.01$ or larger
than $0.1$ would only be explain by
EFCII with NO neutrino spectrum.\\

Summarising, the study of these three ratios can provide relevant
information if values for these ratios are found to be larger than
$0.1$ ($10$) for $R^{\mu\to e\gamma}_{\tau\to \mu\gamma}$ and
$R^{\tau\to e\gamma}_{\tau\to \mu\gamma}$ (for $R^{\mu\to e\gamma}_{\tau\to e\gamma}$)
or smaller than $0.004$ for $R^{\mu\to e\gamma}_{\tau\to \mu\gamma}$,
$0.01$ for $R^{\tau\to e\gamma}_{\tau\to \mu\gamma}$, and
$0.04$ for $R^{\mu\to e\gamma}_{\tau\to e\gamma}$:
such values can be explained only in the case of EFCII with NO spectrum.
If large values for $R^{\mu\to e\gamma}_{\tau\to \mu\gamma}$ and
$R^{\tau\to e\gamma}_{\tau\to \mu\gamma}$ are found, then this would point
to a  relatively
small value for the lightest neutrino mass, smaller than $0.008\eV$;
this should occur consistently with a value for $R^{\mu\to e\gamma}_{\tau\to e\gamma}$ between $0.1$ and $10$.
If instead, $R^{\mu\to e\gamma}_{\tau\to e\gamma}$ is found to be much larger
than $10$, this would imply masses for the lightest neutrino between $0.008\eV$ and $0.04\eV$; consistently, $R^{\mu\to e\gamma}_{\tau\to \mu\gamma}$ and $R^{\tau\to e\gamma}_{\tau\to \mu\gamma}$ should remain smaller than $1$.
Finally, if no signals are seen in all the three ratios and bounds
of $0.004$ ($0.01$) [$0.04$] or smaller can be obtained
for $R^{\mu\to e\gamma}_{\tau\to \mu\gamma}$ ($R^{\tau\to e\gamma}_{\tau\to \mu\gamma}$)
[$R^{\mu\to e\gamma}_{\tau\to e\gamma}$], then this would be consistent with
masses between $0.01\eV$ and $0.02\eV$ for the lightest neutrino,
or otherwise MLFV cannot explain this feature.
On the other hand, all the three MLFV versions, for both the mass orderings,
can explain values for these ratios inside the regions aforementioned,
generally between $0.01$ and $0.1$: this case would be the less favourable for distinguishing the different setups.

These results are generically in agreement with previous analyses performed
in Refs.~\cite{Cirigliano:2005ck,Cirigliano:2006su,Davidson:2006bd,Alonso:2011jd}
and the differences are due to the update input data used here.

\subsubsection*{\boldmath$  B^\text{A}_{\mu\to e}$}

As shown in Eq.~(\ref{BRatioConv}), the ratio of the two branching
ratios with the best present sensitivities is independent from
$\Delta$ and can be used to obtain information about the chirality of
the operators contributing to the $\mu\to e$ conversion process.
On the other hand, if the observation (or non-observation) of the
leptonic radiative rare decays allows to identify the MLFV realisation
from Figs.~\ref{fig:Bmuetaumumlact}, \ref{fig:Bmuetauemlall} and
\ref{fig:Btaumutaueml}, the branching ratio of the $\mu\to e$ conversion in
nuclei could provide the missing information necessary to fix the LFV scale.
As an example, one can assume that an upper bound on
$R^{\mu\to e\gamma}_{\tau\to \mu\gamma}$ of about $0.004$ has been set,
that could be explained by EFCII with a NO neutrino spectrum and a mass
of the lightest neutrino of about $0.014\eV$.
The upper bound on $B^\text{Au}_{\mu\to e}$ implies the upper
bound $v^2/(\muLN\LLF)<5.7\times 10^{-17}$. By fixing the LNV scale to
its lower bound, one finds that these observables can provide information on
the LFV scale that should be larger than about $2\times 10^6\GeV$.
The future expected sensitivity on $B^\text{Al}_{\mu\to e}$ is better than the
presently achieved one by four orders of magnitude. A negative results
of the planned future searches for  $\mu\to e$ conversion
would imply a bound on the LFV scale of about $10^7\GeV$.

\boldmath
\section[\boldmath $b\to s$ Anomalies]{$b\to s$ Anomalies}
\label{Sect:bAnomalies}
\unboldmath

The effective Lagrangian in Eq.~(\ref{EffLagLFV}) contains the operators which provide the most relevant contributions to the $b\to s$ anomalies under discussion\footnote{The complete effective Lagrangian that describes effects in $B$ physics can be found in Ref.~\cite{Alonso:2014csa}. In particular, another operator, with respect to the reduced list in Eq.~(\ref{HEEffOperators}), would contribute at tree level to $C_9$, $\ov e_R\gamma^\mu e_R\ov q_L\gamma_\mu q_L$: this contribution is however negligible for the observables discussed here~\cite{Hiller:2014yaa,Alonso:2015sja}, and then this operator is not considered in the present discussion.}: they are $\O^{(3)}_{LL}$ and $\O^{(5)}_{LL}$, which contribute at tree level to the Wilson coefficients $C_9$ and $C_{10}$ defined in Eq.~(\ref{btosHamiltonian}), satisfying to $\delta C_{10}=-\delta C_9$.

Focussing on the flavour structure of $\O^{(3)}_{LL}$ and $\O^{(5)}_{LL}$, the two operators are invariant under the MFV flavour symmetry $\G_Q\times \G_L$, but can only describe flavour conserving observables which predict universality conservation in both the quark and lepton sectors. In order to describe a process with quark flavour change, it is then necessary to insert powers of the quark Yukawa spurion $\Y_u$. The dominant contributions would arise contracting the flavour indices of the quark bilinear with $\Y_u\Y_u^\dag$: once the spurions acquire their background values, the $b\to s$ transitions are weighted by the $V_{tb}V_{ts}^\ast$ factor appearing in Eq.~(\ref{btosHamiltonian}). Notice that, as $(Y_{u})_{33}=y_t\approx1$, an additional insertion of $\Y_u\Y_u^\dag$ is not negligible and modifies the dominant contributions by $(1+ y_t^{2})$ factors. Further insertions of $\Y_u\Y_u^\dag$ turn out to be unphysical, as they can be written as combinations of the linear and quadratic terms through the Cayley-Hamilton theorem. The complete spurion insertions in $\O^{(3,5)}_{LL}$ can then be written as $\zeta_1\Y_u\Y_u^\dag+\zeta_2(\Y_u\Y_u^\dag)^2$, with $\zeta_{1,2}$ arbitrary coefficients, reflecting the independence of each insertion: the net contribution to the operator is then given by $V_{tb}V_{ts}^\ast(\zeta_1y_t^{2}+\zeta_2y_t^{4})$.

The anomalies in the angular observable $P'_5$ of $B\to K^\ast\mu^+\mu^-$, in the ratios $R_K$ and $R_{K^\ast}$, and in the Branching Ratio of $B_s\to\phi\mu^+\mu^-$ are linked to the possible violation of leptonic universality. NP contributions leading to these effects can be described in terms of insertions of spurion combinations transforming under $\bf 8$ of $SU(3)_{\ell_L}$. The simplest structure is $\Y_e\Y_e^\dag$ that, in the basis defined in Eq.~(\ref{BackgroudSpurionsMFC}), is diagonal and therefore cannot lead to lepton flavour changing transitions. The phenomenological analysis associated to the insertion of this spurionic combination has been performed in Ref.~\cite{Alonso:2015sja}, where the focus was in understanding the consequences of having a setup where lepton universality is violated but lepton flavour is conserved. In Ref.~\cite{Alonso:2015sja}, the Abelian factors in Eq.~(\ref{FullGL}) are considered as active factors of the flavour symmetry and this leads to background values for $\Y_e$, whose largest eigenvalue is of order $1$. It should be noticed that strong constraints on this setup arise when considering radiative electroweak corrections as discussed in Ref.~\cite{Feruglio:2016gvd,Feruglio:2017rjo}.

Focussing only on the non-Abelian factors, as in the tradicional MLFV, the largest entry of $Y_e$ is of the order of $0.01$, as can be seen from Eq.~(\ref{BackgroudSpurionsMFC}). In this scenario, the insertion of $\Y_e$ is subdominant with respect to the insertion of the neutrino spurions: the most relevant are $\g_\nu^\dag\, \g_\nu$ in the MFC, $\Y_\nu\Y_\nu^\dag$ in the EFCI and $\Y_N^\dag \Y_N$ in the EFCII. Once the spurions acquire background values, these contributions reduce to the $\Delta$ characteristic of each case. Similarly to what discussed above for $Y_u$, if the largest eigenvalue of $\Delta$ is of order 1, then additional insertions of the neutrino spurions need to be taken into consideration. The specific contribution depends on the model considered and only a generic form $\sum^2_{n=0} \xi_n \Delta^n$ can be generically written, where $\xi_n$ are arbitrary Lagrangian coefficients, and where the sum is stopped at $n=2$ due to the Cayley-Hamilton theorem.

In Ref.~\cite{Lee:2015qra} the EFCI context has been considered and several processes have been studied, discussing the viability of this version of MLFV to consistently describe the $b\to s$ anomalies.

The aim of this section is to critically revisit the analysis of Ref.~\cite{Lee:2015qra}, and to investigate the other two versions of MLFV. As already mentioned, EFCI will be disfavoured if the Dirac CP violation in the leptonic sector is confirmed, and therefore the viability of MFC and EFCII to describe the $b\to s$ anomalies, consistently with the other (un)observed flavour processes in the $B$ sector, becomes an interesting issue. Moreover, the results obtained in the previous section will be explicitly considered.

\boldmath
\subsection{$B$ Semi-Leptonic Decays}
\unboldmath

In order to facilitate the comparison with Ref.~\cite{Lee:2015qra} similar assumptions will be taken. First of all, setting $C_{10}^\text{SM}=-C_9^\text{SM}$ and considering that the contributions from $\O_{LL}^{(3,5)}$ satisfy to $\delta C_{10}=-\delta C_9$, one can consider a single Wilson coefficient in Eq.~(\ref{btosHamiltonian}): for definiteness, $C_9$ will be retain in what follows. A second relevant assumption is on the matching between the effective operators of the high-energy Lagrangian defined at $\Lambda_\text{LFV}$, Eq.~(\ref{HEEffOperators}), and the low-energy phenomenological description in Eq.~(\ref{btosHamiltonian}): only the tree level relations will be considered in the following, while effects from loop-contributions and from the electroweak running will be neglected. The latter has been recently shown in Ref.~\cite{Feruglio:2016gvd,Feruglio:2017rjo} to lead to a rich phenomenology, especially in EWPO and $\tau$ sector.

Considering explicitly the contributions from $\O_{LL}^{(3,5)}$, and specifying the flavour indexes, one can write
\be
\delta C_{9,\ell\ell'}=\dfrac{\pi}{\alpha_\text{em}}\dfrac{v^2}{\LLF^2}\left(c_{LL,\ell\ell'}^{(3)}+c_{LL,\ell\ell'}^{(5)}\right)\,,
\label{Genericcee}
\ee
where $c_{LL,\ell\ell'}^{(i)}$ can be written in a notation that makes explicit the dependence on the neutrino spurion background\footnote{In Ref.~\cite{Lee:2015qra} a slightly different notation has been adopted, where
\be
c_{LL,\ell\ell'}^{(i)}=\dfrac{\alpha_\text{em}}{\pi}\dfrac{\LLF^2}{v^2}\left[\tilde\xi^{(i)}_0\delta_{\ell\ell'}+\tilde\xi^{(i)}_1\Delta_{\ell\ell'}+\tilde\xi^{(i)}_2\Delta_{\ell\ell'}\right]\,,
\ee
with
\be
\tilde \xi^{(i)}_{j}=\dfrac{\pi}{\sqrt2\alpha_\text{em}G_F\LLF^2} (\zeta^{(i)}_1y_t^2+\zeta^{(i)}_2y_t^4)\xi^{(i)}_i\,.
\ee
}:
\be
c_{LL,\ell\ell'}^{(i)}=\left(\zeta^{(i)}_1y_t^2+\zeta^{(i)}_2y_t^4\right)\left(\xi^{(i)}_0\delta_{\ell\ell'}+\xi^{(i)}_1\Delta_{\ell\ell'}+\xi^{(i)}_2\Delta_{\ell\ell'}\right)\,.
\label{Specificcee}
\ee

In order to explain lepton universality violation, the contributions proportional to $\xi^{(i)}_1$, $\xi^{(i)}_2$, etc. should be at least comparable with $\xi^{(i)}_0$. Consequently, this requires $\Delta_{\ell\ell}\sim 1$, and this allows to fix the scale of LNV: indeed, the bounds in Eqs.~(\ref{LNVScaleBoundMFC}), (\ref{LNVScaleBoundEFCI}) and (\ref{LNVScaleBoundEFCII}) become equalities,
\be
\begin{cases}
\LLN=6\times 10^{14}\GeV\,,\qquad\qquad & \text{for MFC}\\
\muLN=6\times 10^{14}\GeV\,,\qquad\qquad & \text{for EFCI and EFCII}\,.
\end{cases}
\ee

The bounds from LFV purely leptonic processes discussed in the previous section allows to translate this result into specific values for the LFV scale: from the bounds on $\mu\to e$ conversion in nuclei, Fig.~\ref{ScalesPlots}, one obtains that
\be
\begin{cases}
\LLF=4.4\times 10^{5}\GeV\,,\qquad\qquad & \text{for MFC}\\
\LLF=2\times 10^{5}\GeV\,,\qquad\qquad & \text{for EFCI}\\
\LLF=10^{5}\GeV\,,\qquad\qquad & \text{for EFCII}\,.
\end{cases}
\ee
With these results at hand, the order of magnitude for $\delta C_9$ turns out to be
\be
\begin{cases}
\delta C_9=1.3\times10^{-4}\,,\qquad\qquad & \text{for MFC}\\
\delta C_9=6.5\times10^{-4}\,,\qquad\qquad & \text{for EFCI}\\
\delta C_9=2.6\times10^{-3}\,,\qquad\qquad & \text{for EFCII}\,,
\end{cases}
\ee
estimating only the pre-factors appearing in Eq.~(\ref{Genericcee}). These values should now be compared with the ones in Eq.~(\ref{ValuesC9C10}), necessary to explain the anomalies in $b\to s$ decays: the version of MLFV that most contributes to the $C_9$ Wilson coefficient is EFCII, but its contributions are two order of magnitudes too small to explain the $B$ anomalies. It would be only by accident that the parameters of order 1 in Eq.~(\ref{Specificcee}) combine together to compensate such suppression, but this would be an extremely tuned situation.

The only conclusion that can be deduced from this analysis is that all the three versions of MLFV cannot explain deviations from the SM predictions in the Wilson coefficient $C_9$ larger than a few per mil, once taking into consideration the bounds from leptonic radiative decays and conversion of muons in nuclei, contrary to what presented in previous literature.

If the anomalies in the $B$ sector will be confirmed, then it will be necessary to extend the MLFV context. Attempts in this directions have already appeared in the literature, although not motivated by the search for an explanation of the $b\to s$ decay anomalies. The flavour symmetry of the M(L)FV is a continuous global symmetry and therefore, once promoting the spurions to dynamical fields,  its spontaneous breaking leads to the arising of Goldstone bosons. Although it would be possible to provide masses for these new states, this would require an explicit breaking of the flavour symmetry. An alternative is to gauge the symmetry~\cite{Grinstein:2010ve,Feldmann:2010yp,Guadagnoli:2011id,Buras:2011zb,Buras:2011wi,Alonso:2016onw}: the would-be-Goldstone bosons would be eaten by flavour gauge bosons that enrich the spectrum. In recent papers~\cite{Alonso:2017bff,Alonso:2017uky}, a specific gauge boson arising from the chosen gauged flavour symmetry has the specific couplings to explain the $b\to s$ anomalies here mentioned.

\section{Conclusions}
\label{Sect:Conc}

The MFV is a framework to describe fermion masses and mixings and to provide at the same time a sort of flavour protection from beyond the Standard Model contributions to flavour processes. The lack of knowledge of the neutrino mass origin reflects in a larger freedom when implementing the MFV ansatz in the lepton sector: three distinct versions of the MLFV have been proposed in the literature. 

In the present paper, an update of the phenomenological analyses on these setups is presented considering the most recent fit on the neutrino oscillation data.  The recent indication of CP violation in the leptonic sector, if confirmed, will disfavour the very popular MLFV version~\cite{Cirigliano:2005ck} called here EFCI, where right-handed neutrinos are assumed to be degenerate at tree level and the flavour symmetry is $SU(3)_{\ell_L}\times SU(3)_{e_R}\times SO(3)_{N_R}\times CP$. 

The study of the predictions within these frameworks for flavour changing processes has been presented, focussing on leptonic radiative rare decays and muon conversion in nuclei, which provide the stringent bounds. A strategy to disentangle between the different MLFV possibilities has been described: in particular, the next future experiments searching for $\mu\to e\gamma$ and $\mu\to e$ conversion in aluminium could have the power to pinpoint the scenario described here as EFCII~\cite{Alonso:2011jd}, characterised by the flavour symmetry $SU(3)_{\ell_L+N_R}\times SU(3)_{e_R}$, if the neutrino mass spectrum is normal ordered.

An interesting question is whether the present anomalies in the semi-leptonic $B$-meson decays can find an explanation within the M(L)FV context. Contrary to what claimed in the literature, such an explanation would require a scale of New Physics that turns out to be excluded once considering purely leptonic 
processes, the limits on the rate of muon conversion in nuclei being the 
most constraining. These anomalies could find a solution extending/modifying the M(L)FV setup, for example, by gauging the flavour symmetry.

\section*{Acknowledgements}

L.M. thanks the department of Physics and Astronomy of
the Universit\`a degli Studi di Padova for the hospitality during
the writing up of this paper and Paride Paradisi for useful comments
on this project and for all the enjoyable discussions during this visit. D.N.D thanks the Department of Physics of the University of Virginia for the hospitality and P.Q. Hung for the exciting discussions and kind helps.\\
D.N.D. acknowledges partial support by the Vietnam National Foundation for Science and Technology Development (NAFOSTED) under the grant 103.01-2014.89, and by the Vietnam Education Foundation (VEF) for the scholarship to work at the Department of Physics of the University of Virginia.
L.M. and S.T.P acknowledge partial financial support by the European Union's
Horizon 2020 research and innovation programme under the
Marie Sklodowska-Curie grant agreements No 690575 and No 674896.
The work of L.M. was supported in part also by ``Spanish Agencia Estatal de Investigaci\'on'' (AEI) and
the EU ``Fondo Europeo de Desarrollo Regional'' (FEDER) through the
project FPA2016-78645-P, and by the Spanish MINECO through the Centro
de excelencia Severo Ochoa Program under grant SEV-2012-0249
and  by the Spanish MINECO through the
``Ram\'on y Cajal'' programme (RYC-2015-17173).
The work of S.T.P. was supported in part by the INFN
program on Theoretical Astroparticle Physics (TASP) and by
the World Premier International Research Center Initiative (WPI
Initiative), MEXT, Japan.


\footnotesize

\providecommand{\href}[2]{#2}\begingroup\raggedright\endgroup

\end{document}